\documentclass[11pt,a4paper]{article}
\usepackage{amssymb}
\usepackage{amsmath}
\usepackage{amsfonts}
\usepackage{bbm}
\usepackage{amsthm}
\usepackage{mathrsfs}
\usepackage{hyperref}
\usepackage{marvosym}
\usepackage{color}
\usepackage[margin=2.5cm]{geometry}
\usepackage[all,cmtip]{xy}
\usepackage[utf8]{inputenc}
\usepackage{graphicx}

\usepackage{upgreek}

\definecolor{darkred}{rgb}{0.8,0.1,0.1}
\hypersetup{
     colorlinks=false,         
     linkcolor=darkred,
     citecolor=blue,
}

\theoremstyle{plain}
\newtheorem{theo}{Theorem}

\theoremstyle{definition}

\newtheorem{rem}{Remark}

\numberwithin{equation}{section}

\def\nn{\nonumber}

\def\bbR{\mathbb{R}}

\def\bbZ{\mathbb{Z}}

\def\Hom{\mathrm{Hom}}

\def\id{\mathrm{id}}
\def\supp{\mathrm{supp}}

\def\dd{\mathrm{d}}
\def\vol{\mathrm{vol}}
\def\sc{\mathrm{sc}}

\def\1{\mathbbm{1}}
\def\oone{\mathbf{1}}
\def\op{\mathrm{op}}

\def\pr{\mathrm{pr}}

\def\CC{\mathsf{C}}
\def\Man{\mathsf{Man}}
\def\Loc{\mathsf{Loc}}
\def\CRing{\mathsf{C}^\infty\mathsf{Ring}}
\def\Set{\mathsf{Set}}
\def\FMan{\mathsf{FMan}}

\def\Sol{\mathfrak{Sol}}
\def\PP{\mathfrak{P}}

\def\sk{\vspace{2mm}}

\title{%
Poisson algebras for non-linear field theories in the Cahiers topos
}

\author{%
Marco Benini$^{1,a}$ and Alexander Schenkel$^{2,3,b}$ \vspace{2mm}\\
{\small $^1$ Institut f\"ur Mathematik, Universit\"at Potsdam,}\\
{\small Karl-Liebknecht-Str.~24-25, 14476 Potsdam, Germany.}\vspace{1.5mm}\\
{\small $^2$ Department of Mathematics, Heriot-Watt University,}\\
{\small Colin Maclaurin Building, Riccarton, Edinburgh EH14 4AS, United Kingdom.}\vspace{0.2mm}\\
{\small \& Maxwell Institute for Mathematical Sciences, Edinburgh, United Kingdom.}\vspace{1.5mm}\\
{\small ${}^3$ School of Mathematical Sciences, University of Nottingham,}\\
{\small University Park, Nottingham NG7 2RD, United Kingdom.}\vspace{2mm}\\
{\footnotesize \texttt{Email:} 
$^a$\texttt{mbenini87@gmail.com}, $^b$\texttt{aschenkel83@gmail.com}}
 }

\date{October 2016\vspace{-3.5mm}}

\begin{document}

\maketitle

\begin{abstract}
\noindent We develop an approach to construct Poisson algebras for non-linear scalar field theories that is based on the Cahiers topos model for synthetic differential geometry. In this framework the solution space of the field equation carries a natural smooth structure and, following Zuckerman's ideas, we can endow it with a presymplectic current. We formulate the Hamiltonian vector field equation in this setting and show that it selects a family of observables which forms a Poisson algebra. Our approach provides a clean splitting between geometric and algebraic aspects of the construction of a Poisson algebra, which are sufficient to guarantee existence, and analytical aspects that are crucial to analyze its properties.
\vspace{-2mm}
\end{abstract}

\paragraph*{Report no.:} EMPG--16--03\vspace{-2.5mm}
 \vspace{-2.5mm}
\vspace{-2.5mm}


\tableofcontents



\section{\label{sec:intro}Introduction and summary}
Classical field theory is the study of solutions to geometric partial differential
equations (PDEs) on manifolds, which are typically equipped with some 
extra structures such as Lorentzian (or Riemannian) metrics and fiber bundles. 
If the PDE of interest arises as the Euler-Lagrange equation
of some local Lagrangian, it is well-known that 
there is a canonical presymplectic form on the 
space of solutions \cite{Zuckerman}, see also \cite{Khavkine:2014kya} for a recent review. 
An interesting task is then to quantize the solution space along 
this presymplectic form and thereby achieve the transition from 
classical to quantum field theory.
\sk

Even though the program sketched above admits this very simple formulation,
it is usually hard to construct examples in a mathematically rigorous fashion.
The main technical problems are:
1.)~The spaces of field configurations and solutions
are typically infinite-dimensional and therefore they are not described by ordinary manifolds.
As a consequence, one has to work in a broader geometric framework which is capable to describe 
such infinite-dimensional spaces.
2.)~The presymplectic form is typically not strictly symplectic,
but at best weakly symplectic. This complicates the transition to a Poisson algebra of functions
on the solution space, which is the starting point for deformation quantization.
3.)~There is currently no systematic way to study the
deformation quantization of Poisson algebras of functions on infinite-dimensional spaces, even though there
are some recent attempts in this direction \cite{Collini}. 
Hence, the transition from classical to quantum field theory is very hard, which is of course
a well-known fact.
\sk

In this paper we focus on problem 1.)\ and 2.)\ and propose a solution 
in terms of synthetic differential geometry 
\cite{Kock,MoerdijkReyes,Lavendhomme}.
The basic idea of synthetic differential geometry
is to introduce a category of ``generalized smooth spaces''
that contains ordinary manifolds and is closed under forming (co)limits and 
exponential objects. Limits and colimits 
include and generalize common geometric 
constructions like forming subspaces, intersections, unions, 
products or quotients, while exponential objects correspond to ``spaces of mappings''.
Many of the above operations in general do not exist
in ordinary approaches to differential geometry\footnote{For example,
intersections of manifolds are typically no longer manifolds.
Moreover, mapping spaces between two manifolds are infinite-dimensional manifolds
and mapping spaces between two infinite-dimensional manifolds cannot be defined in general, 
see e.g.\ \cite{KrieglMichor}.},
while they always make sense in synthetic differential geometry. 
This flexibility constitutes an evident advantage of the category of ``generalized smooth spaces'' 
that is relevant to synthetic differential geometry. 
This category is also required to contain certain ``infinitesimal spaces'', 
which allow for an intrinsic definition of differential geometric constructions
such as the formation of tangent bundles without going through 
limiting procedures. 
It is worth to explain in non-technical terms how the framework
of synthetic differential geometry allows us to solve problems 1.)\ and 2.)\ above:
Concerning problem 1.), the key point is that a typical field configuration space
is (a subspace of) a mapping space between manifolds,
hence it exists as a generalized smooth space.
A solution space is the subspace of all field configurations satisfying a (possibly non-linear)
field equation and thus is a generalized smooth space as well. The synthetic framework
is therefore flexible enough to do geometry on the spaces relevant for classical field theory.
Concerning problem 2.), the key point is that we can use infinitesimal spaces to obtain
natural definitions of tangent vectors, vector fields and differential forms on the generalized 
smooth spaces appearing in classical field theory. This allows us to study the (pre)symplectic
geometry of solution spaces and to formulate a natural Hamiltonian vector field equation. 
The space of solutions of the Hamiltonian vector field equation is again a generalized 
smooth space and we show that it carries a natural Poisson algebra structure, even in the
case where the presymplectic form is degenerate.
\sk

In order to simplify our presentation, we shall 
use the Cahiers topos \cite{Dubuc} as a 
(well-adapted) model for synthetic differential geometry. 
The choice of a well-adapted model ensures that the standard constructions 
allowed in ordinary differential geometry 
(e.g.\ formation of tangent bundles and transversal intersections) 
are faithfully reproduced by synthetic differential geometry. 
We shall focus on a class of examples of non-linear classical field theories, namely
real scalar fields on Lorentzian manifolds with PDE given by the 
sum of the d'Alembert operator and a (possibly non-polynomial) interaction term.
This includes $\Phi^4$-theory as well as the sine-Gordon model.
Nevertheless, the approach we propose can be vastly generalized to more complicated non-linear field theories, 
e.g.\ those formulated in terms of sections of generic vector bundles 
or even in terms of maps between smooth manifolds, such as the wave map equation ($\sigma$-model). 
We decided to stick to the case of scalar field theory 
in order not to obscure the construction of a Poisson algebra 
with a more involved geometric structure on the field theoretic side. 
We shall also avoid using abstract arguments based on internal topos logic
and often write out our constructions in more elementary terms.
On the one hand, this will simplify the comparison with other approaches
to classical field theory and, on the other hand, it will make our paper better
accessible to readers without any background on topos theory.
\sk

It is important to mention that our construction 
of Poisson algebras for non-linear classical field theories 
does not rely on PDE-analytical properties of the field equation 
or its linearization. The construction we perform holds internally to the Cahiers topos 
without any further requirement and independently of any analytical property of the field equation at hand. 
Only when one wants to study properties of the resulting Poisson algebra in more detail, 
a good control of the Cauchy problem for the field equation or its linearization becomes crucial.
We see this as an advantage compared to other recent approaches \cite{Brunetti:2012ar},
where analytic, geometric and algebraic techniques have to be mixed to construct 
Poisson algebras. One can say that our synthetic approach
introduces a clean splitting between 
abstract geometric/algebraic constructions, which are enough to construct Poisson algebras, 
and PDE-analytical  considerations, which are necessary afterwards for analyzing additional properties. 
See our discussion in Section \ref{sec:conclusion} for more details on this point.
Another advantage of our synthetic approach to classical field theory 
is that it is a suitable starting point for generalizations to gauge theories.
In particular, the groupoids of gauge field configurations appearing in 
our recently proposed homotopy theoretic approach to gauge theories 
\cite{Benini:2015hta} can be easily promoted to groupoid objects
in the Cahiers topos, i.e.\ ``generalized smooth groupoids''.
The relevant homotopy theoretical concepts used in \cite{Benini:2015hta}
generalize to such ``generalized smooth groupoids'' \cite{Groupoids}, while 
locally-convex Lie groupoids (which arise in the framework of
\cite{Brunetti:2012ar}) are not suitable for homotopy theory.
\sk

The outline of the remainder of this paper is as follows:
In Section \ref{sec:prelim} we give a gentle introduction to the Cahiers topos 
and synthetic differential geometry. 
In Section \ref{sec:confspace} we analyze the synthetic geometry of the configuration space
of a scalar field theory and in particular compute its tangent bundle.
The synthetic geometry of the solution space of non-linear scalar field equations on Lorentzian manifolds
is studied in Section \ref{sec:fieldequation}. In Section \ref{sec:Zuckerman}
we formalize the relevant techniques of \cite{Zuckerman} within our framework
and in particular construct a presymplectic current on the solution space.
Our main results are presented in Sections \ref{sec:compact} and \ref{sec:noncompact},
where we construct Poisson algebras for our class of non-linear classical 
field theories by solving suitable Hamiltonian vector field equations. 
Section \ref{sec:conclusion} contains some concluding remarks on the Cauchy problem
within our framework, which we believe to be a good tool for proving additional properties
of our Poisson algebras, e.g.\ the validity of the classical versions
of the axioms of locally covariant quantum field theory  \cite{Brunetti:2001dx}.
Appendix \ref{app:veryexplicit} provides some technical details
of constructions which are used in the main text.


\section{\label{sec:prelim}Preliminaries}
In this section we provide a gentle introduction to the Cahiers topos model
for synthetic differential geometry \cite{Dubuc}, see also \cite{Kock,MoerdijkReyes}.
The Cahiers topos is a category of ``generalized smooth spaces'' 
that exhibits good categorical properties, e.g.\ existence of 
(co)limits and exponential objects, and also contains ``infinitesimal spaces''
which allow for an intrinsic definition of many differential geometric constructions, 
e.g.\ the formation of tangent bundles.
We do not assume the reader to be familiar with the 
theory of (pre)sheaves. All necessary standard concepts 
(e.g.\ Yoneda embedding, Yoneda Lemma and the functor of points perspective)
will be explained explicitly to the extent needed for understanding our constructions by
using our particular example of (pre)sheaf category.
For readers who are familiar with (pre)sheaves and 
synthetic differential geometry this section should 
serve to fix our notations.

\paragraph{Definition of the Cahiers topos:}
The building blocks for the spaces in the Cahiers topos $\CC$ are (finite-dimensional and paracompact) 
manifolds $N$ and infinitesimal spaces $\ell W$ given by the locus of a Weil algebra $W$ over $\bbR$.
Recall that a Weil algebra $W$ is a unital and commutative algebra over $\bbR$ with the following
three properties: 1.)~$W$ is local with maximal ideal $I$ and $W/I \simeq \bbR$.
2.)~$W$ is finite dimensional as a vector space. 3.)~$I$ is nilpotent, i.e.\ there exists
$n\geq 1$ such that $I^n=0$. It follows that $W = \bbR \oplus I$, so
any element $w\in W$ admits a decomposition
$w = \underline{w} + \widehat{w}$, where $\underline{w} \in\bbR$ is the scalar prefactor 
of the unit and $\widehat{w}\in I$ is nilpotent.
An important example of a Weil algebra is the algebra of dual numbers 
$\bbR[\epsilon] := \bbR \oplus \epsilon \bbR$ with product
given by $(a +\epsilon\, b)\, (a' + \epsilon\, b') = a\,a' +\epsilon\, 
(a \,b' + b\, a')$, i.e.\ $\epsilon^2=0$. We follow the standard notations of synthetic differential geometry
and denote the locus of $\bbR[\epsilon]$ by $D := \ell \bbR[\epsilon]$. Loosely speaking,
the infinitesimal space $D$ is an infinitesimally short line, so short that all smooth functions
on $D$ (which are described by $\bbR[\epsilon]$) are fully determined by their first-order Taylor expansion (given by
$a,b\in\bbR$).
\sk

Spaces of the form $N\times \ell W$ are called formal manifolds
and we denote the category of such spaces by $\FMan$. To give a precise definition
of the category $\FMan$, we need some basic terminology from $C^\infty$-rings, 
see e.g.\ \cite{MoerdijkReyes,Joyce}.
A $C^\infty$-ring is a set $A$ together with maps
\begin{flalign}
A_f : A^n = \underbrace{A\times \cdots \times A}_{\text{$n$-times}} \longrightarrow A^m= \underbrace{A\times \cdots \times A}_{\text{$m$-times}}~,
\end{flalign}
for all smooth maps $f : \bbR^n\to \bbR^m$ and $n,m\geq 0$.
These maps must satisfy the following conditions:
1.)~For any $f : \bbR^n \to \bbR^m$ and $g : \bbR^m\to \bbR^l$ smooth,
$A_{g \circ f} = A_g\circ A_f$. 
2.)~For any $n\geq 0$, $A_{\id_{\bbR^n}} = \id_{A^n}$.
3.)~For any projection $\pi_i : \bbR^n \to \bbR$, where $1\leq i\leq n$ and $n\geq 1$,
$A_{\pi_i} = \pi_i : A^n \to A\,,~(a_1,\dots,a_n)\mapsto a_i$.
A morphism between two $C^\infty$-rings
is a map (of sets) $\kappa : A\to B$ such that the diagram
\begin{flalign}
\xymatrix{
\ar[d]_-{A_f} A^n \ar[rr]^-{\kappa^n} && B^n\ar[d]^-{B_f}\\
A^m \ar[rr]_-{ \kappa^m} && B^m
}
\end{flalign}
commutes, for all smooth maps $f : \bbR^n\to \bbR^m$.\footnote{\label{footnote:Cinftyringfunctor}Notice
that this definition of $C^\infty$-ring is equivalent to saying that a $C^\infty$-ring
is a finite-product preserving functor $\mathcal{A} : \mathsf{Cart}\to \Set$ from Cartesian spaces to sets.
(More precisely, $\mathsf{Cart}$ is the category with objects given by all Cartesian spaces $\bbR^k$, $k\in\bbZ_{\geq 0}$,
and morphisms given by all smooth maps between Cartesian spaces.)
The set $A$ in our first definition is obtained by evaluating this functor on the one-dimensional Cartesian space 
$\bbR^1$, i.e.\ $A= \mathcal{A}(\bbR^1)$.
In this picture, a morphism of $C^\infty$-rings is simply a natural transformation between 
finite-product preserving functors.
}
We denote the category of $C^\infty$-rings by $\CRing$.
Notice that any $C^\infty$-ring $A$ is in particular a unital and 
commutative algebra over $\bbR$; all algebra operations can be realized 
as polynomial mappings $f : \bbR^n \to \bbR^m$ (which are smooth maps), e.g.\
the product $\mu_A : A\times A \to A$ is given by $\mu_A = A_{\mu_{\bbR}}$, where
$\mu_{\bbR} : \bbR\times \bbR\to \bbR\,,~(c,c')\mapsto c\,c'$ is the product on $\bbR$,
and the unit element $\eta_A : \{\ast\} \to A$ is given by $\eta_A = A_{\eta_{\bbR}}$,
where $\eta_{\bbR} : \{\ast\} \to \bbR\,,~\ast\mapsto 1$ is the unit in $\bbR$. 
Moreover, scalar multiplication on $A$ by $\lambda\in\bbR$ is given by $A_{\lambda} : A\to A$,
where $\lambda : \bbR \to\bbR\,,~c\mapsto \lambda\,c$.
\sk

We give some important examples of $C^\infty$-rings which will play a major role in our work:
Given any manifold $N$, the set of smooth functions $C^\infty(N)$ from $N$ to $\bbR$ is
a $C^\infty$-ring with
\begin{flalign}
C^\infty(N)_f : C^\infty(N)^n\longrightarrow C^\infty(N)^m~,~~(h_1,\dots,h_n)\longmapsto f\circ (h_1,\dots,h_n)~,
\end{flalign}
for all $f: \bbR^n\to\bbR^m$ smooth. Regarding Cartesian spaces $\bbR^k$ as manifolds,
we obtain that $C^\infty(\bbR^k)$ are $C^\infty$-rings. Even more, $C^\infty(\bbR^k)$ are the free $C^\infty$-rings
with $k$ generators, i.e.\ $\Hom_{\CRing} (C^\infty(\bbR^k),A)\simeq A^k$ for any other $C^\infty$-ring $A$.
Given any Weil algebra $W$, then there exists a unique $C^\infty$-ring structure on $W$ which extends its algebra structure,
see e.g.\ \cite[Proposition 1.5]{Dubuc} or \cite[Theorem III.5.3]{Kock}. Explicitly,
\begin{flalign}
W_f : W^n \longrightarrow W^m~,~~(w_1,\dots,w_n)\longmapsto  f(w_1,\dots,w_n)~,
\end{flalign}
for all $f: \bbR^n\to\bbR^m$ smooth, where the right-hand-side is understood in terms of Taylor expansion 
of $f$ in all nilpotents $ \widehat {w_i}$ at the point $(\underline{w_0},\dots,\underline{w_n})\in\bbR^n$.
(Notice that the Taylor expansion in nilpotents terminates at some finite order). 
\sk

With these preparations we can now give a precise definition of the category $\FMan$.
It is the opposite of the full subcategory of $\CRing$ with objects given by
$C^\infty(N)\otimes_{\infty} W$, where $N$ is any 
(finite-dimensional and paracompact) manifold and $W$ is any Weil algebra.
Here $\otimes_{\infty}$ denotes the coproduct in $\CRing$. 
In order to simplify notations, we shall also denote formal manifolds by symbols like
$t = N\times\ell W$ and $t^\prime = N^\prime\times \ell W^\prime$.
By definition, the morphisms in $\FMan$ are given by 
\begin{flalign}\label{eqn:FManmorphi}
\Hom_{\FMan} (t, t^\prime) 
:= \Hom_{\CRing}\big(C^\infty(N^\prime)\otimes_{\infty} W^\prime ,C^\infty(N)\otimes_{\infty} W \big)~.
\end{flalign}
We can equip the category $\FMan$ with a Grothendieck topology by declaring a covering family
to be a family of morphisms of the form
\begin{flalign}
\Big\{ \xymatrix{U_i \times \ell W \ar[rr]^-{\rho_i \times \id_{\ell W}} & & N \times \ell W}\Big\} ~,
\end{flalign}
where $\{ U_i \stackrel{\rho_i}{\longrightarrow} N\}$  is an ordinary open cover
of the manifold $N$. The Cahiers topos is then by definition the category of sheaves on this site, i.e.\
\begin{flalign}
\CC := \mathrm{Sh}\big(\FMan\big)~.
\end{flalign}
Objects in $\CC$ are sheaves, i.e.\ functors $X : \FMan^\op \to \Set$ to the category of sets (also called presheaves)
which satisfy the sheaf condition with respect to the notion of covering described above.
Morphisms $f : X \to Y$ in $\CC$ are natural transformations between such functors.

\paragraph{Embedding of manifolds into the Cahiers topos:}
The Cahiers topos $\CC$ is a  category of generalized smooth spaces that includes,
as we shall see later, various kinds of infinite-dimensional spaces.
Moreover, $\CC$ also contains (in a suitable way to be specified below) various objects 
which describe well-known spaces such as manifolds and infinitesimal spaces.
The key point is that the Yoneda embedding allows us to embed formal manifolds into the Cahiers topos.
Explicitly, given any object $t=N\times \ell W$ in $\FMan$, 
its Yoneda embedding $\iota(t): \FMan^\op \to \Set$ is the $\Set$-valued presheaf on $\FMan$ 
that acts on objects $t^\prime=N^\prime\times \ell W^\prime$ as
\begin{subequations}\label{eqn:Yoneda}
\begin{flalign}
\iota(t)(t^\prime) := 
\Hom_{\FMan}(t^\prime, t) 
= \Hom_{\CRing}\big(C^\infty(N)\otimes_{\infty} W, C^\infty(N^\prime)\otimes_{\infty} W^\prime\big)~,
\end{flalign}
and on morphisms $f: t^\prime \to t^{\prime\prime}$ as
\begin{flalign}
 \iota(t)(f)  : \Hom_{\FMan}(t^{\prime\prime}, t) \longrightarrow \Hom_{\FMan}(t^\prime, t)~,~~
\big(g  : t^{\prime\prime}\to t\big) \longmapsto \big(g \circ f : t^{\prime}\to t\big)~.
\end{flalign}
\end{subequations}
The fact that the presheaf $\iota(t)$ defined above is actually a sheaf on $\FMan$, 
i.e.\ that the site is subcanonical, is due to \cite{Dubuc}. 
As a consequence, the (presheaf) Yoneda embedding $t\mapsto \iota(t)$ factors through $\CC$ 
and defines a functor $\iota: \FMan \to \CC$ from formal manifolds to the Cahiers topos,
which we also call Yoneda embedding. 
This functor is fully faithful, i.e.\ the set of morphisms (in $\FMan$)
between two objects $t$ and $t^\prime$
is isomorphic to the set of morphisms (in $\CC$)
between $\iota(t)$ and $\iota(t^\prime)$.
Loosely speaking, this means that the theory of formal manifolds together with their morphisms
can be equivalently described within the Cahiers topos $\CC$.
The interpretation of the sheaf $\iota(t)$ defined in \eqref{eqn:Yoneda}
is that of the {\em functor of points} of the formal manifold $t=N\times \ell W$.
Again loosely speaking, \eqref{eqn:Yoneda} tells us all possible ways in which any other formal manifold
$t^\prime$ maps smoothly into $t$ and this is enough information
to know everything about the smooth structure on $t$. 
A similar interpretation is used for generic objects $X$ in $\CC$: The sets
$X(t^\prime)$ obtained by evaluating the functor
$X : \FMan^\op\to \Set$ on $t^\prime$ tell us all possible ways in which
$t^\prime$ is mapped smoothly to the generalized smooth space $X$.
This is formalized by Yoneda's Lemma, which states that there is an isomorphism
\begin{flalign}
\Hom_{\CC}\big(\iota(t^\prime), X\big) \simeq X(t^\prime)~,
\end{flalign}
for any object $t^\prime$  in $\FMan$. Hence, it is justified to call
elements in $X(t^\prime)$ generalized points (of type $t^\prime$) of $X$.
\sk

Regarding manifolds $N$ as formal manifolds of the form $N\times\ell \bbR$,
the Yoneda embedding restricts to a fully faithful embedding $\iota : \Man\to\CC$
of the category of (finite-dimensional and paracompact) manifolds $\Man$
into the Cahiers topos. The same holds true for infinitesimal spaces $\ell W$,
which can be regarded as formal manifolds of the form $\{\ast\} \times \ell W$,
where $\{\ast\}$ is any one-point manifold.
To simplify notation, we shall drop the embeddings $\iota$ and simply write
$N$, $\ell W$ and $t= N\times \ell W$ for the objects in $\CC$ 
which are given by embedding manifolds, infinitesimal spaces and formal manifolds.

\paragraph{Categorical properties of the Cahiers topos:}
As any category of sheaves (technically called a 
Grothendieck topos), the Cahiers topos $\CC$ has good categorical properties,
see e.g.\ \cite[Chapter III]{MacLaneMoerdijk}.
All (small) limits and colimits exist in $\CC$ and the former can be computed object-wise 
(i.e.\ like in the category of presheaves). 
Computing colimits in $\CC$ is more complicated as one first forms the colimit in the category of presheaves 
(which is computed object-wise)
and then applies the sheafification functor to the result.
Special instances of limits, which will be of major importance below,
are products: Given two objects $X$ and $Y$ in $\CC$, their product $X\times Y$ in $\CC$
is the sheaf specified by the functor $X\times Y : \FMan^\op \to \Set$ that acts on objects as
\begin{flalign}
(X\times Y)(t) := X(t) \times Y (t)~,
\end{flalign}
where on the right-hand-side $\times$ denotes the Cartesian product in $\Set$.
Notice also that $\CC$ has a terminal object $\{\ast\}$ which is the sheaf specified by
the functor $\{\ast\} : \FMan^\op \to \Set$ that acts on objects as
\begin{flalign}
\{\ast\}(t) := \{\ast\}~,
\end{flalign}
where on the right-hand-side $\{\ast\}$ denotes the terminal object in $\Set$, i.e.\ a singleton.
The embedding $\iota : \FMan \to \CC$
of formal manifolds into the Cahiers topos preserves the terminal object 
and products (and also transversal pullbacks of manifolds).
\sk

Another good categorical property of $\CC$ is the existence of exponential 
objects (also called mapping spaces): Given two objects $X$ and $Y$ in $\CC$,
the exponential object $Y^X$ in $\CC$ (interpreted as the object of mappings from $X$ to $Y$)
is the sheaf specified by the functor $Y^X : \FMan^\op\to\Set$ that acts on objects as
\begin{flalign}\label{eqn:exponentialobject}
Y^X(t) := \Hom_{\CC}(t\times X , Y)~,
\end{flalign}
where on the right-hand-side $t$ is interpreted as an object in $\CC$ via the Yoneda embedding.
We recall that $(-)^X : \CC \to \CC$ and $Y^{(-)} : \CC^\op \to\CC$ are functors:
Explicitly, given any morphism $f : Y\to Z$ in $\CC$, then $f^X : Y^X \to Z^X$ is the morphism
in $\CC$ which is specified by the natural transformation with components
\begin{flalign}
 \resizebox{0.91\textwidth}{!}{$f^X : \Hom_{\CC}(t\times X, Y) \longrightarrow\Hom_{\CC}(t\times X, Z)~,~~
 \big(h: t \times X\to  Y \big) \longmapsto  \big(f \circ h: t \times X\to  Z \big) ~.$}\label{eqn:fpush}
\end{flalign}
Similarly, given any morphism $g : X\to Z$ in $\CC$, then $Y^g : Y^Z \to Y^X$  is the morphism
in $\CC$ which is specified by the natural transformation with components
\begin{flalign}
 \resizebox{0.91\textwidth}{!}{$Y^g : \Hom_{\CC}(t\times Z, Y) \longrightarrow \Hom_{\CC}(t\times X, Y)~,~~
 \big(h: t\times  Z \to  Y \big) \longmapsto  \big(h\circ (\id_{t} \times g) : t\times X\to  Y \big) ~.$}\label{eqn:gpull}
\end{flalign}
Finally, as in any category admitting finite products and exponential objects (i.e.\ a Cartesian closed category),
there exist natural isomorphisms
\begin{flalign}\label{eqn:exponentialproperties}
\{\ast\}^X \simeq \{\ast\}~,\quad X^{\{\ast\}} \simeq X~,\quad  (Y\times Z)^{X}\simeq Y^X\times Z^X~,\quad
X^{Y\times Z}\simeq (X^Y )^Z~,
\end{flalign}
for all objects $X,Y,Z$ in $\CC$.

\paragraph{Basic aspects of synthetic differential geometry:}
By the Yoneda embedding $\iota : \FMan \to \CC$,
we can regard the infinitesimal spaces $\ell W \simeq \{\ast \} \times \ell W$ 
as objects in $\CC$, i.e.\ as generalized smooth spaces. An important example
of such an infinitesimal space is $D = \ell \bbR[\epsilon]$, which, as we have argued above,
should be interpreted as an infinitesimally short line.
More precisely, the Cahiers topos has a line object $R := \iota(\bbR)$ which is given by embedding (via Yoneda)
the real line $\bbR$ into $\CC$ and there is a monomorphism $D\to R$ in $\CC$; explicitly,
$D\to R$ is given by the $\CRing$-morphism $C^\infty(\bbR) \to \bbR[\epsilon]$ which Taylor expands
a function $h\in C^\infty(\bbR)$ to first order around $0$, i.e.\ 
$h\mapsto h(0)  + \epsilon\, h^\prime(0)$.
Moreover, $D$ contains the zero element, which is the point
$0 : \{\ast\}\to D$ specified by the $\CRing$-morphism
$\bbR[\epsilon] \to \bbR\,,~a + \epsilon \,b \mapsto a$.
\sk

Using the object $D$, we can define (the total space of) the tangent bundle
of {\em any} object $X$ in $\CC$ in terms of the exponential object
$TX := X^D$. One should think of $TX$ as the space of infinitesimally short curves in $X$,
which contain the information of a base point (the image of the zero element) and
a tangent vector at this base point (the direction of the curve).
The projection  $\pi : TX\to X$ is given by 
exponentiation with the zero element $0 : \{\ast\}\to D$, i.e.\
\begin{flalign}
\xymatrix{
TX  = X^D \ar[rr]^-{\pi:= X^0} && X^{\{\ast\}} \simeq X~.
}
\end{flalign}
Because $(-)^D : \CC\to \CC$ is a functor,
the assignment of the total spaces of the tangent bundles is functorial. Moreover, 
from \eqref{eqn:fpush} and \eqref{eqn:gpull} it follows that 
for any morphism $f :X\to Y$ in $\CC$ the diagram
\begin{flalign}\label{eqn:tangentfunctor}
\xymatrix{
\ar[d]_\pi TX = X^D \ar[rr]^-{Tf := f^D} && Y^D = TY\ar[d]^-\pi\\
X\ar[rr]_-{f} && Y
}
\end{flalign}
commutes.
An important result \cite[Proposition II.1.12]{MoerdijkReyes}
is that the tangent bundles defined as above coincide in the case of finite-dimensional 
manifolds with the ordinary tangent bundles: Explicitly, given any object
$N$ in $\Man$, then $T\iota(N) = \iota(N)^D \simeq \iota(TN)$, where 
on the right-hand-side $TN $ denotes the ordinary tangent bundle of $N$.
In particular, our convention to suppress  all $\iota$
is consistent with the formation of tangent bundles  $TN$.
\sk

The synthetic construction of tangent bundles is just one example of
how infinitesimal spaces can be used to simplify  constructions in
differential geometry and also to generalize them to the generalized smooth spaces 
described by the Cahiers topos. For a more complete presentation
of the synthetic approach to differential geometry we refer the reader to 
the standard textbook references \cite{Kock,MoerdijkReyes,Lavendhomme}.


\section{\label{sec:confspace}Configuration space of a scalar field theory}
Let $M$ be a finite-dimensional manifold which we interpret as spacetime. 
The field configurations of a real scalar field on $M$ are given by all smooth mappings
$\Phi : M\to \bbR$ from $M$ to the real numbers, i.e.\ by the {\em set} $C^\infty(M)$.
Making use of the Cahiers topos, we can define a generalized smooth space
of scalar field configurations by considering the exponential object 
$R^M$ in $\CC$, where $M$ is regarded as an object in $\CC$ via the Yoneda embedding. 
The advantage of the object $R^M$ in $\CC$ compared to the set $C^\infty(M)$
is that $R^M$ is a generalized smooth space, hence we can do synthetic differential
geometry on it. The functor $R^M : \FMan^\op \to\Set$ describing the object $R^M$ in $\CC$
has the following more elementary description:
Given any object $t=N\times \ell W$ in $\FMan$, we have isomorphisms
\begin{flalign}
\nn R^M(t) &= \Hom_{\CC}(t \times M , R)\\
\nn &= \Hom_{\FMan}( t \times M,\bbR )\\
\nn &= \Hom_{\CRing}\big(C^\infty(\bbR) , C^\infty(N\times M)\otimes_{\infty} W\big)\\
\nn &\simeq C^\infty(N\times M)\otimes_{\infty} W \\
&\simeq C^\infty(N\times M)\otimes_{\bbR} W~.\label{eqn:confexplicit}
\end{flalign}
In the first step we have used the definition of exponential objects \eqref{eqn:exponentialobject}
and in the second step that both objects $t\times M = (N\times M)\times\ell W$ and $R$
in $\CC$ are representable, i.e.\ obtained by the fully faithful Yoneda embedding $\iota: \FMan\to \CC $.
Step three is simply the definition of morphisms in $\FMan$, see \eqref{eqn:FManmorphi}, and
step four uses that $C^\infty(\bbR)$ is the free $C^\infty$-ring with one generator.
The last isomorphism is due to the fact that the coproduct in $\CRing$ with a Weil algebra is 
isomorphic to the coproduct of algebras over $\bbR$ (cf.\ \cite[Theorem III.5.3]{Kock}). 
Hence, generalized points (of type $t=N\times\ell W$) of $R^M$ are given
by elements $\Phi\in C^\infty(N\times M)\otimes_{\bbR} W$, i.e.\ they are
Weil algebra $W$-valued fields on the product $N \times M$ of spacetime $M$ and a manifold $N$.
Notice that global points, i.e.\ morphisms $\Phi : \{\ast\} \to R^M$ in $\CC$
or equivalently elements of $R^M(\{\ast\})$, are simply given by ordinary scalar fields $\Phi \in C^\infty(M)$.
Using \eqref{eqn:gpull}, we obtain that the configuration spaces are functorial, i.e.\
$R^{(-)} : \Man^\op \to \CC$.
\sk

We now compute the tangent bundle of our configuration spaces $R^M$.
Using the definition from the previous section, the total space
of the tangent bundle of $R^M$ is given by $T R^M := (R^M)^D$, with $D = \ell \bbR[\epsilon]$
the infinitesimally short line. Using \eqref{eqn:exponentialproperties}, there exist isomorphisms
\begin{flalign}\label{eqn:tangconfisos}
T R^M  = (R^M)^D \simeq R^{M\times D} \simeq (R^D)^M\simeq (T\bbR)^M~,
\end{flalign}
i.e.\ the total space of the tangent bundle of the space of $\bbR$-valued fields is the
space of $T\bbR$-valued fields. Similar to \eqref{eqn:confexplicit}, we have a more elementary description
of the functor $T R^M : \FMan^\op \to\Set$, which is given by
\begin{flalign}
T R^M(t)  \simeq \Hom_{\CC} (t \times M \times D,R)
 \simeq C^\infty(N\times M)\otimes_{\bbR} W \otimes_{\bbR} \bbR[\epsilon]~,\label{eqn:Tconfexplicit}
\end{flalign}
for any object $t$ in $\FMan$. Hence, a generalized point of $T R^M$ is  
given by an element $C^\infty(N\times M)\otimes_{\bbR} W \otimes_{\bbR} \bbR[\epsilon]$,
which we can write as $\Phi + \epsilon\, \Psi$, where $\Phi,\Psi\in C^\infty(N\times M)\otimes_{\bbR} W$
are generalized points of $R^M$. The role of $\Phi$ is that of a base point and $\Psi$  is a tangent vector at $\Phi$.
\sk

We finish this section by noting that $R^M$ is a $C^\infty$-ring object in $\CC$.\footnote{Following the point of view
of footnote \ref{footnote:Cinftyringfunctor}, a $C^\infty$-ring object in
$\CC$ is by definition a finite-product preserving functor $\mathcal{A} : \mathsf{Cart} \to \CC$.} This
will be used in the next section in order to define (possibly non-polynomial) interaction
terms in field equations on $R^M$, e.g.\ the sine-Gordon term. 
The fact that $R^M$ is a $C^\infty$-ring object in $\CC$ follows immediately 
once one notices that both the Yoneda embedding $\iota: \FMan \to \CC$ 
and the functor $(-)^M: \CC \to \CC$ preserve products. 
Nevertheless, we provide explicit formulas for this $C^\infty$-ring structure
as they will be needed later to write out the non-linear field equations.
As already mentioned above, the line object $R= \iota(\bbR)$ in $\CC$ is a $C^\infty$-ring
object in $\CC$ because the Yoneda embedding $\iota : \FMan \to \CC$ preserves 
products and $\bbR$ is a $C^\infty$-ring (valued in sets). 
We denote the $\CC$-morphisms corresponding to smooth maps $f: \bbR^n\to \bbR^m $
by $R_f : R^n\to R^m$. 
Given $f:  \bbR^n\to \bbR^m $ smooth, we define a $\CC$-morphism
\begin{flalign}
\xymatrix{
(R^M)^n  \simeq (\underbrace{R\times \cdots \times R}_{\text{$n$-times}})^M
\ar[rrr]^-{(R^M)_f := {R_f}^M}&&&  (\underbrace{R\times \cdots \times R}_{\text{$m$-times}})^M \simeq (R^M)^m
}~,
\end{flalign}
where we have used the fact that $(-)^M : \CC\to \CC$
is a functor which preserves products, see  \eqref{eqn:exponentialproperties}. 
This structures $R^M$ as a $C^\infty$-ring object in $\CC$.
For later use, we shall also provide an explicit formula for $(R^M)_\rho : R^M\to R^M$ 
in the case where $\rho : \bbR \to \bbR$ is a smooth map between one-dimensional Cartesian spaces.
Using \eqref{eqn:fpush}, we obtain that $(R^M)_\rho : R^M\to R^M$  is the natural transformation
with components
\begin{flalign}
 \resizebox{0.91\textwidth}{!}{$(R^M)_\rho : \Hom_{\CC} (t \times M , R) \longrightarrow \Hom_{\CC} (t \times M , R)~,~~
\big( h : t \times M \to R \big)\longmapsto \big( R_\rho\circ h : t \times M \to R\big)~,$}\label{eqn:CinftyringRupM}
\end{flalign}
for all objects $t$ in $\FMan$.
Using the elementary description of $R^M(t)$ given in \eqref{eqn:confexplicit},
we can further simplify \eqref{eqn:CinftyringRupM} as
\begin{flalign}
(R^M)_\rho : C^\infty(N\times M) \otimes_{\bbR} W \longrightarrow C^\infty(N\times M) \otimes_{\bbR} W~,~~\Phi 
\longmapsto \rho\circ \Phi~,\label{eqn:CinftyringRupMexplicit}
\end{flalign}
where the right-hand-side is understood in terms of Taylor expansion of $\rho$ in the nilpotent terms of $\Phi$.
Explicitly, we can write $\Phi = \underline{\Phi} + \widehat{\Phi}$, where $\underline{\Phi}\in C^\infty(N\times M)$
is the prefactor of the unit in $W$ and $\widehat{\Phi}\in C^\infty(N\times M) \otimes_{\bbR} I$ is nilpotent,
and \eqref{eqn:CinftyringRupMexplicit} is given by expanding at each point
$(x,p)\in N\times M $ the expression $\rho\big(\underline{\Phi} (x,p) + \widehat{\Phi} (x,p)\big)$
in the nilpotent $\widehat{\Phi} (x,p)$ around the point $\underline{\Phi} (x,p) \in \bbR$.


\section{\label{sec:fieldequation}Field equation and solution space}
In this section we study dynamical aspects of a class of non-linear scalar field theories.
Let $M$ be an oriented and time-oriented globally hyperbolic Lorentzian
manifold. The d'Alembert operator on the configuration space $R^M$
is the $\CC$-morphism $\square_{M}^{} : R^M \to R^M$ given by the natural transformation
with components
\begin{flalign}\label{eqn:dAlembert}
\square_{M}^{} := \square^\mathrm{vert}_{M} \otimes_{\bbR} \id_W : C^\infty(N\times M)\otimes_{\bbR} 
W \longrightarrow C^\infty(N\times M)\otimes_{\bbR} W~,
\end{flalign}
where we have used the elementary description of $R^M(t)$ given in \eqref{eqn:confexplicit}. 
Moreover, the vertical d'Alembert operator 
$\square^\mathrm{vert}_{M}: C^\infty(N\times M) \to C^\infty(N\times M)$ 
is defined with respect to the vertical Lorentzian geometry of the trivial bundle $N\times M \to N$ 
(in particular, it involves only derivatives along $M$). 
More explicitly, \eqref{eqn:dAlembert} lifts the ordinary d'Alembert operator $\square_M^{}$ from $C^\infty(M)$ to 
$C^\infty(N\times M)\otimes_{\bbR} W$ in the following way: Choosing a basis $\{e_i\}$ of $W$, 
we can expand each $\Phi \in C^\infty(N\times M)\otimes_{\bbR} W$ as
$\Phi = \sum_i \Phi^i\,e_i$, where $\Phi^{i}\in C^\infty(N\times M)$,
and $\square_{M}^{}$ acts on $\Phi $ by acting with $\square^\mathrm{vert}_{M}$
on each component $\Phi^i$ without mixing them.
\sk

Given now any smooth map $\rho : \bbR\to \bbR$, we add \eqref{eqn:dAlembert}
and \eqref{eqn:CinftyringRupMexplicit} to obtain an equation of motion
operator $P_M^{} : R^M\to R^M$ on the configuration space $R^M$ 
which is in general non-linear. (In the following 
we shall keep $\rho$ fixed and suppress it from the notations.) 
Explicitly, the $\CC$-morphism $P_{M}^{} : R^M \to R^M$ is the natural transformation
with components
\begin{flalign}\label{eqn:EOMoperator}
P_M^{} : C^\infty(N\times M)\otimes_{\bbR} W \longrightarrow C^\infty(N\times M)\otimes_{\bbR} W~,~~
\Phi \longmapsto \square_M^{}\Phi + \rho\circ \Phi~.
\end{flalign}
For example, we could choose in \eqref{eqn:EOMoperator}
the function $\rho : \bbR\to \bbR\,,~x\mapsto \lambda \, x^3$, for some coupling constant 
$\lambda\in \bbR$, to obtain the equation of motion of $\Phi^4$-theory. 
As another example, we could choose $\rho : \bbR\to \bbR\,,~x\mapsto \sin x$
to obtain the sine-Gordon equation.
\sk

The space of solutions to the equation of motion \eqref{eqn:EOMoperator} 
is constructed by the pullback
\begin{flalign}\label{eqn:solutionspace}
\xymatrix{
\ar@{-->}[d]  \Sol(M) \ar@{-->}[rr] && R^M\ar[d]^-{P_M^{}}\\
\{\ast\} \simeq \{\ast\}^M \ar[rr]_-{0^M} && R^M
}
\end{flalign}
in $\CC$. Because pullbacks (as special kinds of limits) in $\CC$ exist, the solution space $\Sol(M)$ to {\em any}
(non-linear) field equation is always a generalized smooth space (i.e.\ an object in $\CC$).
This is a clear advantage of the synthetic framework over ordinary approaches
to infinite-dimensional differential geometry, such as locally-convex manifolds, 
where the spaces of solutions to non-linear field equations in general do not carry a natural smooth structure.
Using the elementary description of $R^M(t)$ given in \eqref{eqn:confexplicit},
we observe that generalized points of $\Sol(M)$ 
are given by elements $\Phi \in C^\infty(N\times M)\otimes_{\bbR} W$
which satisfy $P_M^{}(\Phi) =0$, i.e.\
\begin{flalign}\label{eqn:Solexplicit}
\Sol(M)(t) \simeq \Big\{\Phi \in C^\infty(N\times M)\otimes_{\bbR} W \, :\,  
P_M^{}(\Phi)  =0 \Big\}~,
\end{flalign}
for all objects  $t$ in $\FMan$.
\sk

The solution spaces are functorial: Let $\Loc$ denote the category with objects 
given by oriented and time-oriented globally hyperbolic Lorentzian
manifolds $M$ (of a fixed dimension, say $m$) and morphisms given by causal 
embeddings $f : M\to M^\prime$.\footnote{A causal embedding 
$f : M\to M^\prime$ is an orientation and time-orientation preserving isometric embedding, 
whose image is open and causally compatible, i.e.\ $J_{M^\prime}^\pm(f(p)) \cap f(M) = f(J_M^\pm(p))$ 
for all $p \in M$. Here $J_M^\pm(p)$ denotes the causal future/past of $p\in M$ 
consisting of all points of $M$ which can be reached by a future/past-directed smooth causal curve 
in $M$ stemming from $p$, see e.g.\ \cite{Bar:2007zz}.
}
Then $R^{(-)} : \Loc^\op \to \CC$ is a functor and the equation of motion operator
\eqref{eqn:EOMoperator} is a natural transformation $P : R^{(-)}\to R^{(-)}$ between 
functors from $\Loc^\op$ to $\CC$. As a consequence, the pullback diagram \eqref{eqn:solutionspace}
which defines the solution spaces is functorial, and by universality of limits
we obtain that the solution spaces are given by a functor 
\begin{flalign}
\Sol : \Loc^\op \longrightarrow \CC~.
\end{flalign}

We now shall compute the tangent bundle of the solution space $\Sol(M)$, for any object $M$ in $\Loc$.
As the tangent functor is given by exponentiation $T (-) = (-)^D : \CC\to \CC $ with the object
$D = \ell \bbR[\epsilon]$, it is a right adjoint functor (of the functor $(-)\times D : \CC\to \CC$)
and as such it preserves limits. In particular, applying the tangent functor to the pullback diagram 
\eqref{eqn:solutionspace}, we obtain that $T\Sol(M) = \Sol(M)^D$ is given by the pullback
\begin{flalign}\label{eqn:Tsolutionspace}
\xymatrix{
\ar@{-->}[d]  T\Sol(M)  \ar@{-->}[rr] && T R^M \ar[d]^-{T P_M^{} }\\
 \{\ast\} \simeq T \{\ast\}^{M} \ar[rr]_-{T 0^{M}} && TR^M
}
\end{flalign}
in $\CC$. Using the elementary description of $T R^M(t)$ given in \eqref{eqn:Tconfexplicit},
we obtain that generalized points of $T\Sol(M)$ are elements $\Phi + \epsilon\,\Psi \in 
C^\infty(N\times M)\otimes_{\bbR} W\otimes_{\bbR} \bbR[\epsilon]$, with $\Phi,\Psi\in 
C^\infty(N\times M)\otimes_{\bbR} W$, which satisfy $P_M^{} (\Phi+ \epsilon\,\Psi) =0$.
Expanding the latter equation in the nilpotent $\epsilon$ (with $\epsilon^2=0$),
we obtain the two equations $P_M^{}(\Phi) =0$ and 
\begin{flalign}\label{eqn:EOMoperatorlin}
P_{M,\,\Phi}^{\mathrm{lin}} \Psi := \square_M^{}\Psi + \big(\rho^\prime\circ \Phi\big)\,\Psi =0~,
\end{flalign}
where $\rho^\prime :\bbR\to \bbR$ is the derivative of $\rho : \bbR\to \bbR$.
Notice that \eqref{eqn:EOMoperatorlin} is the linearization of the equation of motion operator \eqref{eqn:EOMoperator}
around the solution $\Phi$, i.e.\ $\Psi$ satisfies a linear equation of motion.
In summary, the functor $T\Sol(M) : \FMan^\op \to\Set$ has an elementary description given by
\begin{flalign}\label{eqn:TSoleasy}
T\Sol(M)(t) \simeq \Big\{\Phi + \epsilon\,\Psi \in
C^\infty(N\times M)\otimes_{\bbR} W\otimes_{\bbR} \bbR[\epsilon]\,:\, P_M^{}(\Phi) =0\, ,~P_{M,\,\Phi}^{\mathrm{lin}} \Psi =0 \Big\}~,
\end{flalign}
for all objects $t$ in $\FMan$.


\section{\label{sec:Zuckerman}Zuckerman's presymplectic current}
In \cite{Zuckerman}, Zuckerman has shown that any field theory specified by a local Lagrangian 
admits an associated presymplectic current specified on an appropriately defined
solution space of the Euler-Lagrange equation.
Notice that our field equation \eqref{eqn:EOMoperator} is the Euler-Lagrange equation
of the scalar field Lagrangian $\mathrm{L}_M^{}$ given by the sum of 
the usual kinetic term $-\frac{1}{2}\dd_M^{} \Phi \wedge \ast_M^{} \dd_M^{}\Phi$
and the potential term $(V\circ \Phi) \, \vol_{M}$, where $V := \int^{\cdot} \rho : \bbR\to \bbR$
is any primitive of the smooth map $\rho : \bbR\to\bbR$ and $\vol_M\in\Omega^m(M)$ 
is the volume form on the oriented Lorentzian manifold $M$ (of dimension $m$). 
Loosely speaking, Zuckerman's presymplectic current is constructed as follows:
One first takes the differential $\dd$ (along the field configuration space) of the Lagrangian
and notices that it can be written as $\dd \mathrm{L}_M^{} = \mathrm{EL}_M^{} + \dd_M^{} \theta_M^{}$,
where $\mathrm{EL}_M^{}$ is the Euler-Lagrange equation, $\dd_M^{}$ is the differential along spacetime
and $\theta_M^{}$ is a $\Omega^{m-1}(M)$-valued $1$-form on the field configuration space.
Pulling back $\theta_M^{}$ to a $\Omega^{m-1}(M)$-valued $1$-form on the solution space,
one obtains a $\Omega^{m-1}(M)$-valued $2$-form on the solution space, 
the presymplectic current, by taking the differential $\mathrm{u}_M^{} = \dd \theta_M^{}$ 
along the solution space. An essential property
of $\mathrm{u}_M^{}$ is that it takes values in the space $\Omega^{m-1}_{\dd}(M)$ 
of closed $m{-}1$-forms on $M$.
We shall now formalize the relevant part of this construction for our model in terms of
the Cahiers topos.

\paragraph{The $1$-form $\theta_M^{}$:}
The $\Omega^{m-1}(M)$-valued $1$-form $\theta_M^{}$ on the solution space
$\Sol(M)$ is given by a $\CC$-morphism
\begin{flalign}\label{eqn:thetaform}
\theta_M^{} : T\Sol(M) \longrightarrow \Omega^{m-1}(M)~,
\end{flalign}
which we shall now describe in some detail.
\sk

The target of \eqref{eqn:thetaform} is the generalized smooth
space of $m{-}1$-forms on $M$, which is the sheaf specified by
the functor $\Omega^{m-1}(M) : \FMan^\op \to \Set$ that acts on objects as
\begin{flalign}\label{eqn:formspace}
\Omega^{m-1}(M)(t) := \Omega^{0,m-1}(N\times M)\otimes_{\bbR} W~,
\end{flalign}
where $\Omega^{0,m-1}(N\times M)$ denotes the vector space 
of $(0,m{-}1)$-forms on the product manifold
$N\times M$ and $\otimes_\bbR$ is the tensor product of real vector spaces.\footnote{The
object $\Omega^{m-1}(M)$ in $\CC$ defined by \eqref{eqn:formspace} can also be
obtained by equipping the usual set of forms $\Omega^{m-1}(M)$ with its canonical convenient vector space structure
(cf.\ \cite{KrieglMichor}) and using the fully faithful embedding $j : \mathsf{ConVec}\to \CC$
of the category of convenient vector spaces into the Cahiers topos, see \cite{KockConvenient} 
and in particular \cite{KockReyes}.
}
(Elements in $\Omega^{0,m-1}(N\times M)$ are differential forms on 
$N\times M$ which are of degree zero in $N$ and of degree $m{-}1$ in $M$, 
see e.g.\ \cite{Zuckerman} for more details on this bi-grading.)
\sk

In the case under analysis, namely the non-linear scalar field, the components of \eqref{eqn:thetaform} are given by
\begin{flalign}\label{eqn:thetaformcomp} 
\theta_M^{} : T\Sol(M)(t) \longrightarrow \Omega^{0,m-1}(N\times M)\otimes_{\bbR} W~,~~
\Phi +\epsilon\,\Psi  \longmapsto - \Psi \ast_M^{}\dd_{M}^{}\Phi~,
\end{flalign}
where the differential
\begin{subequations}
\begin{flalign}
\dd_{M}^{} :=\dd_{M}^{\mathrm{vert}}\otimes_{\bbR}\id_{W} : \Omega^{0,p}(N\times M)\otimes_{\bbR}W
\longrightarrow \Omega^{0,p+1}(N\times M)\otimes_{\bbR}W
\end{flalign}
and the Hodge operator
\begin{flalign}
\ast_M^{} := \ast_M^{\mathrm{vert}} \otimes_{\bbR}\id_{W} : \Omega^{0,p}(N \times M)\otimes_{\bbR}  W 
\longrightarrow \Omega^{0,m-p}(N\times M)\otimes_{\bbR}  W
\end{flalign}
\end{subequations}
are defined in analogy to \eqref{eqn:dAlembert}.
From the definition \eqref{eqn:thetaformcomp}, 
we observe that
\eqref{eqn:thetaform} is $R$-linear 
with respect to the fiber $R$-module structure on $T\Sol(M)$
 given by the $\CC$-morphisms 
\begin{subequations}\label{eqn:tangentVecSpstructure}
\begin{flalign}
+ : T\Sol(M) \times_{\Sol(M)}T\Sol(M) \longrightarrow T\Sol(M)~,\quad
\cdot : R\times T\Sol(M)\longrightarrow T\Sol(M)~,
\end{flalign}
with components
\begin{flalign}
+ : \big(\Phi + \epsilon\,\Psi_1,\Phi + \epsilon\,\Psi_2\big)\longmapsto \big(\Phi + \epsilon\, (\Psi_1+\Psi_2)\big)~,\quad
\cdot : \big(c, \Phi + \epsilon\,\Psi\big)\longmapsto \big(\Phi + \epsilon\,c\,\Psi\big)~.
\end{flalign}
\end{subequations}
The term $c\,\Psi$ appearing in the definition of $\cdot$ is the multiplication
of $\Psi \in C^\infty(N\times M)\otimes_{\bbR} W$
by $c\in R(t) \simeq C^\infty(N)\otimes_{\bbR} W$, which is
regarded as an element in $C^\infty(N\times M)\otimes_{\bbR} W$ that is constant along $M$.
The fiber product $T\Sol(M) \times_{\Sol(M)}T\Sol(M) $ is defined as usual by
the pullback diagram
\begin{flalign}
\xymatrix{
\ar@{-->}[d] T\Sol(M) \times_{\Sol(M)}T\Sol(M) \ar@{-->}[rr] && T\Sol(M)\ar[d]^-{\pi}\\
T\Sol(M)\ar[rr]_-{\pi} && \Sol(M)
}
\end{flalign}
in $\CC$.

\paragraph{Vector fields on $\Sol(M)$:}
The space of vector fields on $\Sol(M)$ is described by the generalized smooth space 
 $\Gamma^\infty (T\Sol(M))$ of sections of the tangent bundle $\pi : T\Sol(M) \to \Sol(M)$, which is
carved out of the exponential object $T\Sol(M)^{\Sol(M)}$ by the following
pullback
\begin{flalign}
\xymatrix{
\ar@{-->}[d]\Gamma^\infty (T\Sol(M)) \ar@{-->}[rr] && T\Sol(M)^{\Sol(M)}\ar[d]^-{\pi^{\Sol(M)}}\\
\{\ast\} \ar[rr]_-{e} && \Sol(M)^{\Sol(M)}
}
\end{flalign}
The $\CC$-morphism 
$e : \{\ast\} \to \Sol(M)^{\Sol(M)}$ is the identity
element in $\Sol(M)^{\Sol(M)}$, which is given explicitly by the components
\begin{flalign}
e : \{\ast\} \longrightarrow \Hom_{\CC}\big(t\times\Sol(M) , \Sol(M)\big) ~,~~\ast \longmapsto \pr_{\Sol(M)}^{}~,
\end{flalign}
where $\pr_{\Sol(M)}^{} : t\times\Sol(M) \to \Sol(M)$ denotes the projection $\CC$-morphism on the 
factor $\Sol(M)$.
\sk

A generalized point of $\Gamma^\infty (T\Sol(M))$ is therefore given by 
a $\CC$-morphism $v: t \times \Sol(M)\to T\Sol(M)$ 
for which the diagram
\begin{flalign}\label{eqn:genvecfield}
\xymatrix{
\ar[rrd]_-{\pr_{\Sol(M)}^{}~~}t \times \Sol(M) \ar[rr]^-{v}&& T\Sol(M)\ar[d]^-{\pi}\\
&& \Sol(M)
}
\end{flalign}
in $\CC$ commutes. 
In other words, $\Gamma^\infty (T\Sol(M))$
is specified by the functor $\Gamma^\infty (T\Sol(M)): \FMan^\op \to \Set$ 
that acts on objects as
\begin{flalign}\label{eqn:vecfieldexplicit}
\Gamma^\infty (T\Sol(M))(t) = 
\Big\{v \in \Hom_{\CC}\big( t\times \Sol(M),T\Sol(M)\big)\,:\,
\pi \circ v =\pr_{\Sol(M)}^{}\Big\}~,
\end{flalign}
for all objects  $t$ in $\FMan$.

\paragraph{The global $1$-form $\theta_M^{}$:}
It will be convenient for our constructions 
to take the global point of view on differential forms on $\Sol(M)$,
see e.g.\ \cite[Chapter 6.1]{Lavendhomme}.
In this perspective, the $\Omega^{m-1}(M)$-valued $1$-form \eqref{eqn:thetaform}
is promoted to a $\CC$-morphism (denoted with abuse of notation by the same symbol)
\begin{flalign}\label{eqn:thetaglobal}
\theta_M^{} : \Gamma^\infty(T\Sol(M))\longrightarrow \Omega^{m-1}(M)^{\Sol(M)}~,
\end{flalign}
which is an assignment of $\Omega^{m-1}(M)$-valued functions on $\Sol(M)$ to vector fields on $\Sol(M)$.
At the level of components, \eqref{eqn:thetaglobal} assigns to each generalized point
of $\Gamma^\infty(T\Sol(M))$,
i.e.\ each $\CC$-morphism $v : t \times \Sol(M)\to T\Sol(M)$
satisfying the section condition $\pi\circ v = \pr_{\Sol(M)}^{}$, the 
$\CC$-morphism 
\begin{flalign}\label{eqn:thetaglobalcomp}
\theta_M^{}(v) :=  \theta^{}_M \circ v :t \times\Sol(M) \longrightarrow \Omega^{m-1}(M)~.
\end{flalign}
Using \eqref{eqn:thetaformcomp}, we observe that 
\eqref{eqn:thetaglobal} is an $R^{\Sol(M)}$-module morphism for the following
`point-wise' $R^{\Sol(M)}$-module structures on $\Gamma^\infty(T\Sol(M))$ and $\Omega^{m-1}(M)^{\Sol(M)}$:
The sum $\CC$-morphisms
\begin{subequations}\label{eqn:Abgroup}
\begin{flalign}
+ : \Gamma^\infty(T\Sol(M))\times \Gamma^\infty(T\Sol(M)) &\longrightarrow \Gamma^\infty(T\Sol(M))~,\\
+ : \Omega^{m-1}(M)^{\Sol(M)}\times \Omega^{m-1}(M)^{\Sol(M)}&\longrightarrow \Omega^{m-1}(M)^{\Sol(M)}~,
\end{flalign}
\end{subequations}
are obtained from \eqref{eqn:tangentVecSpstructure} via \eqref{eqn:fpush}. Explicitly, they
are specified on generalized points $v,v^\prime : t\times \Sol(M) \to T\Sol(M)$, satisfying the section condition
\eqref{eqn:genvecfield}, and $\omega,\omega^\prime : t \times\Sol(M)\to \Omega^{m-1}(M)$ 
by the commutative diagrams
\begin{flalign}
\xymatrix@C=0.05em{
 \ar[drr]_-{(v,v^\prime)~~~} t \times \Sol(M)\ar[rr]^-{v+v^\prime} && T\Sol(M)  && \ar[drr]_-{(\omega,\omega^\prime)~~~} t \times \Sol(M) \ar[rr]^-{\omega+\omega^\prime} && \Omega^{m-1}(M) \\
&& T\Sol(M) \times_{\Sol(M)}T\Sol(M) \ar[u]_-{+}&& && \Omega^{m-1}(M)\times\Omega^{m-1}(M)\ar[u]_-{+} 
}
\end{flalign}
where we have used the notation $(f , g) : X\to Y\times Z$ 
for the unique morphism defined out of $f : X\to Y$ and $g : X\to Z$ 
and universality of the product.
Similarly, the $R^{\Sol(M)}$-action $\CC$-morphisms
\begin{subequations}\label{eqn:RSolmodule}
\begin{flalign}
\cdot : R^{\Sol(M)} \times \Gamma^\infty(T\Sol(M)) &\longrightarrow \Gamma^\infty(T\Sol(M))~,\\
\cdot : R^{\Sol(M)} \times \Omega^{m-1}(M)^{\Sol(M)} &\longrightarrow \Omega^{m-1}(M)^{\Sol(M)}~,
\end{flalign}
\end{subequations}
are obtained from \eqref{eqn:tangentVecSpstructure} via \eqref{eqn:fpush}.
Explicitly, for generalized points $F : t \times\Sol(M)\to R$, 
$v : t \times\Sol(M) \to T\Sol(M)$ 
and $\omega : t \times\Sol(M)\to \Omega^{m-1}(M)$, 
the above-mentioned $R^{\Sol(M)}$-action $\CC$-morphisms are specified by the commutative diagrams
\begin{flalign}
\xymatrix@C=0.5em{
 \ar[drr]_-{(F,v)~~~} t\times \Sol(M)\ar[rr]^-{F\cdot v} && T\Sol(M)  && \ar[drr]_-{(F,\omega)~~~} t \times \Sol(M) \ar[rr]^-{F\cdot \omega} && \Omega^{m-1}(M) \\
&& R\times T\Sol(M) \ar[u]_-{\cdot}&& && R\times\Omega^{m-1}(M)\ar[u]_-{\cdot} 
}
\end{flalign}
In the terminology of \cite[Chapter 6.1.2]{Lavendhomme}, this shows that
$\theta_M^{}$ given in \eqref{eqn:thetaglobal} is a (classical) 
$\Omega^{m-1}(M)$-valued global $1$-form on $\Sol(M)$.

\paragraph{The presymplectic current $\mathrm{u}_M^{}$:}
The presymplectic current $\mathrm{u}_M^{}$ is by definition the 
exterior derivative (along the solution space) of $\theta_M^{}$. 
Explicitly, $\mathrm{u}_M^{}$ is the $\CC$-morphism
\begin{flalign}
\mathrm{u}_M^{}:= \dd\theta_M^{} : \Gamma^\infty(T\Sol(M))\times \Gamma^{\infty}(T\Sol(M)) \longrightarrow \Omega^{m-1}(M)^{\Sol(M)}
\end{flalign}
that is defined on generalized points $v,v^\prime : t \times \Sol(M)\to T\Sol(M)$, satisfying
the section condition \eqref{eqn:genvecfield}, by Koszul's formula
\begin{flalign}\label{eqn:presympcurrent}
\mathrm{u}_M^{}(v,v^\prime) = v\big(\theta_M^{}(v^\prime)\big) - v^\prime\big(\theta_M^{}(v)\big) - \theta_M^{}\big([v,v^\prime]\big)~,
\end{flalign}
where the first two terms involve the action 
of $v$ and respectively $v^\prime$ on $\Omega^{m-1}(M)^{\Sol(M)}$ in terms of 
directional derivatives and the third term involves the Lie bracket on $\Gamma^\infty(T\Sol(M))$,
see \cite[Chapter 6.1.2]{Lavendhomme} and Appendix \ref{app:veryexplicit} for details.
By construction, $\mathrm{u}_M^{} = \dd \theta_M^{}$ is an exact 
global $2$-form on $\Sol(M)$
and hence in particular closed, i.e.\ $\dd \mathrm{u}_M^{} =0$.
An explicit expression for $\mathrm{u}_M^{}(v,v^\prime)$ is derived 
in Appendix \ref{app:veryexplicit} (see in particular \eqref{eqn:presympcurrentultraexplicit}), 
from which we observe that $\mathrm{u}_M^{}$ is a $\CC$-morphism to the generalized smooth space
$\Omega^{m-1}_\dd(M)^{\Sol(M)}$, i.e.\ 
\begin{flalign}\label{eqn:currentOmegadd}
\mathrm{u}_M^{}  : \Gamma^\infty(T\Sol(M))\times \Gamma^{\infty}(T\Sol(M)) \longrightarrow \Omega_{\dd}^{m-1}(M)^{\Sol(M)}~.
\end{flalign}
Here $\Omega^{m-1}_{\dd}(M)$ denotes the subsheaf of $\Omega^{m-1}(M) : \FMan^\op \to \Set$
that is specified by
\begin{flalign}\label{eqn:closedform}
\Omega^{m-1}_\dd(M) (t) := \mathrm{Ker}\Big(\dd_M^{} : \Omega^{0,m-1}(N\times M)\otimes_{\bbR}W \to \Omega^{0,m}(N\times M)\otimes_{\bbR}W\Big)~,
\end{flalign}
for all objects $t$ in $\FMan$.


\section{\label{sec:compact}Poisson algebra for compact Cauchy surfaces}
In this section we assume that $M$ is an $m$-dimensional oriented and 
time-oriented globally hyperbolic Lorentzian manifold (i.e.\ an object in $\Loc$)
which admits a compact Cauchy surface $\Sigma \hookrightarrow M$. This will simplify the construction
of a Poisson algebra of observables for the non-linear field theory specified by the field equation
\eqref{eqn:EOMoperator} and its corresponding presymplectic current \eqref{eqn:currentOmegadd}.
The case of not necessarily compact Cauchy surfaces requires some additional care and will be discussed in Section
\ref{sec:noncompact}.
\sk

Our strategy is as follows: We introduce an integration $\CC$-morphism 
$\int_{\Sigma} : \Omega^{m-1}(M) \to R$, which, after composition with the Zuckerman current 
\eqref{eqn:currentOmegadd}, defines a presymplectic form $\omega_M^{}$ (i.e.\ a closed
$R$-valued global $2$-form) on $\Sol(M)$.
By considering a suitable pullback diagram in the Cahiers topos 
$\CC$, we construct a generalized smooth space describing those pairs $(F,v)$ 
of smooth functions $F$ and vector fields $v$ on $\Sol(M)$ 
which satisfy the Hamiltonian vector field equation. Loosely speaking, the latter equation is 
given by $\dd F = \omega_{M}^{}(v,-)$.
It is important to stress that in general there exist $F$ which do not admit a
Hamiltonian vector field, because $\omega_{M}^{}$ is {\em not} strictly non-degenerate.
Hence, the Hamiltonian vector field equation selects a suitable 
class of smooth functions $F$, which we may call {\em admissible}. 
We will then show that the space of pairs $(F,v)$ consisting of
admissible functions $F$ and their Hamiltonian vector fields $v$
can be  equipped with a Poisson algebra structure.

\paragraph{Presymplectic form:} 
In analogy to \eqref{eqn:dAlembert}, we define a $\CC$-morphism
$\int_{\Sigma} : \Omega^{m-1}(M)\to R$ 
by setting for its components
\begin{flalign}\label{eqn:integration}
\int_{\Sigma} := \int_{\Sigma}^{\mathrm{vert}}\otimes_{\bbR} \id_W : \Omega^{0,m-1}(N\times M)\otimes_{\bbR}W
\longrightarrow C^\infty(N)\otimes_{\bbR} W~,
\end{flalign}
where $\int_{\Sigma}^{\mathrm{vert}}$ is the vertical integration on the trivial fibration
$N\times M\to N$. Notice that this is where the requirement of a compact Cauchy surface enters.
The restriction of the integration morphism to closed forms, i.e.\ $\int_{\Sigma} : \Omega_{\dd}^{m-1}(M)\to R$,
just depends on the homology class $[\Sigma]\in \mathrm{H}_{m-1}(M)$ and hence it is independent
of the choice of Cauchy surface. As a consequence, composing the Zuckerman current 
\eqref{eqn:currentOmegadd} with the integration morphism ${\int_{\Sigma}}^{\Sol(M)} : \Omega^{m-1}_{\dd}(M)^{\Sol(M)}
\to R^{\Sol(M)} $ defines a $\CC$-morphism
\begin{flalign}\label{eqn:symplectic}
\omega_{M}^{} := {\int_{\Sigma}}^{\Sol(M)} \, \circ\,  \mathrm{u}_M^{} :  \Gamma^\infty(T\Sol(M))\times \Gamma^\infty(T\Sol(M)) 
\longrightarrow R^{\Sol(M)}~,
\end{flalign}
which does not depend on the choice of Cauchy surface.
Because $\mathrm{u}_M^{}$ is a closed $\Omega^{m-1}_{\dd}(M)$-valued
$2$-form on $\Sol(M)$ and integration is $R$-linear, it follows that $\omega_M^{}$ 
is a closed global $2$-form on $\Sol(M)$, i.e.\ a presymplectic form.

\paragraph{Hamiltonian vector field equation:}
We can now formalize the Hamiltonian vector field equation, which 
at the level of generalized points $F : t \to R^{\Sol(M)}$
and $v : t \to \Gamma^\infty(T\Sol(M))$ is given by
\begin{flalign}\label{eqn:hamVFequationexplicit}
\dd F = \iota_{v} (\omega_M^{})~,
\end{flalign}
as an equation in $\Omega^1(\Sol(M))(t)$.
Here $\dd$ denotes the differential and $\iota_v( \omega_M^{})$ the interior product,
see \cite[Chapter 6.1.2]{Lavendhomme} for more details.
The generalized space $\PP(M)$ of solutions to the equation \eqref{eqn:hamVFequationexplicit}
is then given by the pullback
\begin{flalign}\label{eqn:hamVFequation}
\xymatrix{
\ar@{-->}[d] \PP(M)\ar@{-->}[rr] && \Gamma^\infty(T\Sol(M))\ar[d]^-{\iota_{(\,\cdot\,)}(\omega_M^{})}\\
R^{\Sol(M)}\ar[rr]_-{\dd}&& \Omega^1(\Sol(M))
}
\end{flalign}
in $\CC$. Explicitly, the generalized smooth space $\PP(M)$ 
is specified by the functor $\PP(M) :\FMan^\op \to \Set$ that acts on objects as
\begin{flalign}
\PP(M)(t) = \Big\{(F,v) \in R^{\Sol(M)}(t)\times \Gamma^\infty(T\Sol(M))(t)  \, :\, \dd F = \iota_v(\omega_M^{})\Big\}~,
\end{flalign}
for all objects $t$ in $\FMan$. We call $F$ an admissible observable
and $v$ a Hamiltonian vector field corresponding to $F$.

\paragraph{Poisson algebra structure on $\PP(M)$:}
As the Hamiltonian vector field equation \eqref{eqn:hamVFequationexplicit}
is $R$-linear, it follows that $\PP(M)$ is an $R$-module. 
The $R$-module structure on 
$\PP(M)$ is inherited from the $R$-module structures 
on $R^{\Sol(M)}$ and $\Gamma^\infty(T\Sol(M))$ 
via universality of the pullback.
\sk

The $R$-module $\PP(M)$ carries an $R$-algebra structure
with product and unit $\CC$-morphisms
\begin{subequations}\label{eqn:PoisRalg}
\begin{flalign}
\cdot : \PP(M)\times\PP(M)\longrightarrow \PP(M)~,\quad \oone : \{\ast\} \longrightarrow \PP(M)~,
\end{flalign}
defined on generalized points by
\begin{flalign}
\cdot : \big((F,v), (F^\prime,v^\prime)\big) \longmapsto \big(F\cdot F^\prime , F\cdot v^\prime + F^\prime\cdot v\big)~,
\quad \oone : \ast\longmapsto (1 , 0)~.
\end{flalign}
\end{subequations}
Here $F\cdot F^\prime$ denotes the product on $R^{\Sol(M)}$,
and $F\cdot v^\prime$ and $F^\prime\cdot v$ the $R^{\Sol(M)}$-module structure
on $\Gamma^{\infty}(T\Sol(M))$, see \eqref{eqn:RSolmodule}.
Using the (graded) Leibniz rule for the differential and $R^{\Sol(M)}$-linearity
of the interior product (see \cite[Chapter 6.1.2]{Lavendhomme}),
we can confirm that the product closes on $\PP(M)$, i.e.\
\begin{flalign}
\dd(F\cdot F^\prime) = F\cdot (\dd F^\prime) +  F^\prime\cdot (\dd F) 
=   F\cdot \iota_{v^\prime}( \omega_M^{})+ F^\prime\cdot \iota_{v}(\omega_M^{} )
= \iota_{ F\cdot v^\prime + F^\prime\cdot v }(\omega_M^{})~.
\end{flalign}
The unit element lies in $\PP(M)$ because $\dd 1 = 0 = \iota_0(\omega_M^{})$.
\sk

Finally, we equip $\PP(M)$ with a Poisson bracket $\CC$-morphism
\begin{subequations}\label{eqn:PoisPalg}
\begin{flalign}
\big\{\, \cdot\, , \,\cdot\,\big\} : \PP(M)\times\PP(M)\longrightarrow \PP(M)~,
\end{flalign}
which is defined on generalized points by
\begin{flalign}
\big\{\, \cdot\, , \,\cdot\,\big\} : \big((F,v), (F^\prime,v^\prime)\big) \longmapsto 
\big(\iota_{v}\iota_{v^\prime}(\omega_{M}^{}), [v,v^\prime]\big)~.
\end{flalign}
\end{subequations}
Using the Cartan calculus properties \cite[Chapter 6.1.2, Proposition 6]{Lavendhomme},
we can confirm that the Poisson bracket closes on $\PP(M)$, i.e.\
\begin{flalign}
\dd \iota_{v}\iota_{v^\prime}(\omega_{M}^{}) = (\mathcal{L}_v - \iota_v\dd)\iota_{v^\prime}(\omega_M^{})
= \iota_{[v,v^\prime]}(\omega_M^{})  + \iota_{v^\prime} \mathcal{L}_v(\omega_M^{}) - \iota_v\mathcal{L}_{v^\prime}(\omega_{M}^{}) = \iota_{[v,v^\prime]}(\omega_M^{})  ~,
\end{flalign}
where $\mathcal{L}_v = \iota_v \, \dd + \dd\, \iota_v$ is the Lie derivative.
In the second equality we have used that $\dd\omega_M^{}=0$ and in the last equality that
$\mathcal{L}_v(\omega_M^{}) = \dd\iota_{v}\omega_M^{} = \dd\dd F =0 $ (and similar for $(F^\prime,v^\prime)$).
The Poisson bracket is clearly antisymmetric and, using again the Cartan calculus,
one easily confirms the derivation property
\begin{flalign}
\big\{(F,v), (F^\prime,v^\prime)\cdot (F^{\prime\prime},v^{\prime\prime})\big\} = 
\big\{(F,v), (F^\prime,v^\prime)\big\} \cdot (F^{\prime\prime},v^{\prime\prime}) + (F^\prime,v^\prime)\cdot 
\big\{(F,v), (F^{\prime\prime},v^{\prime\prime})\big\}
\end{flalign}
and the Jacobi identity
\begin{flalign}
\big\{(F,v), \big\{ (F^\prime,v^\prime),  (F^{\prime\prime},v^{\prime\prime})\big\}\big\}  + \mathrm{cycl} = 0~.
\end{flalign}
In these calculations one also has to use that the Lie bracket $[\,\cdot\,,\,\cdot\,]$
on $\Gamma^\infty(T\Sol(M))$ satisfies
the derivation property $[v,F\cdot v^\prime] = F\cdot [v,v^\prime] + \mathcal{L}_v(F)\cdot v^\prime $ 
and the Jacobi identity, see e.g.\ \cite[Chapter 3]{Lavendhomme}.
Summing up, we have obtained
\begin{theo}\label{theo:Poissoncompact}
Let $M$ be an object in $\Loc$ which admits a compact Cauchy surface
$\Sigma\hookrightarrow M$. Then $\PP(M)$, defined as the pullback in \eqref{eqn:hamVFequation},
is a Poisson algebra object in the Cahiers topos $\CC$ when equipped with the
$R$-module structure inherited via pullback, the $R$-algebra structure \eqref{eqn:PoisRalg}
and the Poisson bracket \eqref{eqn:PoisPalg}.
\end{theo}


\section{\label{sec:noncompact}Poisson algebra for arbitrary Cauchy surfaces}
If $M$ is any object in $\Loc$, i.e.\ an $m$-dimensional oriented and time-oriented globally hyperbolic
Lorentzian manifold with not necessarily compact Cauchy surfaces $\Sigma\hookrightarrow M$,
then the integration of the presymplectic current in \eqref{eqn:symplectic} is ill-defined.
We shall resolve this issue by restricting the tangent bundle $\pi : T\Sol(M)\to\Sol(M)$
to what we call the ``spacelike compact tangent bundle'' $\pi : T_{\mathrm{sc}}\Sol(M)\to \Sol(M)$.
Loosely speaking, the fibers of $T_{\mathrm{sc}}\Sol(M)$ will be
the solutions of the linearized equation of motion \eqref{eqn:EOMoperatorlin}
that are of spacelike compact support.
The restriction of the presymplectic current 
to spacelike compact vector fields $\Gamma^\infty(T_{\sc}\Sol(M))$ takes values in
the generalized smooth space $\Omega^{m-1}_{\mathrm{sc},\,\dd}(M)^{\Sol(M)}$ 
of functions on $\Sol(M)$ with values in closed $m{-}1$-forms on $M$ with spacelike compact support.
As a consequence, the restriction of $\mathrm{u}_M^{}$ to spacelike compact vector fields 
(at least in one argument) can be integrated over a not necessarily compact Cauchy surface, 
and we can formalize the Hamiltonian vector field equation for arbitrary objects in $\Loc$.
Using similar arguments as in Section \ref{sec:compact}, this will lead to a Poisson algebra 
for all objects $M$ in $\Loc$.

\paragraph{The spacelike compact tangent bundle:}
Let $M$ be any object in $\Loc$ and let us
denote by $\mathcal{K}_M^{}$ the directed set of compact subsets $K\subseteq M$
with preorder relation given by subset inclusion $K\subseteq K^\prime$.
For any $K\in \mathcal{K}_M^{}$, 
the open submanifold $M\setminus J_M^{}(K)$ of $M$ 
is an object in $\Loc$, where $J_M^{}(K) := J_M^+ (K)\cup J_M^-(K)$ 
is the union of the causal future and past of $K$.
We denote the canonical $\Loc$-morphism by 
$j : M\setminus J_M^{}(K)\to M$. By functoriality, $j$ induces a $\CC$-morphism
$\Sol(j) : \Sol(M)\to \Sol(M\setminus J_M^{}(K))$ and hence by
functoriality of the tangent bundle a $\CC$-morphism
\begin{flalign}
T\Sol(j) : T\Sol(M)\longrightarrow T\Sol(M\setminus J_M^{}(K))~.
\end{flalign}
We define the generalized smooth space $T_{J_M^{}(K)}\Sol(M)$
by the pullback
\begin{flalign}
\xymatrix@C=5em{
\ar@{-->}[d]T_{J_M^{}(K)}\Sol(M) \ar@{-->}[rr]&& T\Sol(M)\ar[d]^-{T\Sol(j)}\\
\Sol(M\setminus J_M^{}(K))\ar[rr]_-{\Sol(M\setminus J_M^{}(K))^{D\to \{\ast\}}}&&T\Sol(M\setminus J_M^{}(K))
}
\end{flalign}
in $\CC$, where the lower horizontal arrow is the zero section.
Using \eqref{eqn:TSoleasy}, generalized points of $T_{J_M^{}(K)}\Sol(M)$
are given by those elements $\Phi + \epsilon\,\Psi\in T\Sol(M)(t)$ 
which satisfy $\supp(\Psi)\subseteq N\times J_M^{}(K)$. (In other words, the restriction
of $\Psi$ to $N\times M\setminus J_M^{}(K)$ is zero.)
The total space of the spacelike compact tangent bundle is defined as the colimit
\begin{flalign}\label{eqn:Tsc}
T_{\sc}\Sol(M) := \mathrm{colim} \big(T_{J_M^{}(-)}\Sol(M) : \mathcal{K}_M\to \CC \big)
\end{flalign}
in $\CC$. 
It is important to recall from Section \ref{sec:prelim} that
colimits in $\CC$ can be computed as the sheafification of the presheaf colimit (i.e.\ object-wise colimit).
As a result, we obtain that the generalized points
of $T_{\sc}\Sol(M)$ are given by those elements $\Phi + \epsilon\,\Psi\in T\Sol(M)(t)$ 
(cf.\ \eqref{eqn:TSoleasy})
such that for any $x\in N$ there exists an open neighborhood $U \subseteq N$ of $x$ 
and a compact subset $K\subseteq M$ with the property that
$\Psi_U \in C^\infty(U \times M) \otimes_{\bbR} W$ (the restriction of $\Psi$ to $U \times M$) 
has support in $U \times J_M^{}(K)$. (It is important that both $K$ and $U$ are allowed to change 
with $x\in N$. In particular, the uniform condition that $\Psi$ has support in some $N\times J_M^{}(K)$, 
which results from the presheaf colimit, does not define a sheaf.)
In other words, $T_{\sc}\Sol(M)$ is the subsheaf
of $T\Sol(M) : \FMan^\op\to\Set$  that is specified by
\begin{multline}\label{eqn:Tscexplicit}
T_{\sc}\Sol(M)(t) 
= \Big\{\Phi+\epsilon\,\Psi \in T\Sol(M)(t): \\
\forall x\in N~ \exists\, U \ni x \mbox{ open}\,,~ K\in \mathcal{K}_M^{}:~ \supp (\Psi_U)\subseteq U \times J_M^{}(K) \Big\}~,
\end{multline}
for all objects $t$ in $\FMan$. Clearly, $T_{\sc}\Sol(M)$ is the total space of a bundle over
$\Sol(M)$ with projection $\CC$-morphism
\begin{flalign}\label{eqn:SCbundle}
\pi : T_{\sc}\Sol(M)\longrightarrow \Sol(M)
\end{flalign}
induced by the tangent bundle $\pi : T\Sol(M)\to \Sol(M)$, i.e.\ $\pi : T_{\sc}\Sol(M)\to \Sol(M)$ is a subbundle
of the tangent bundle. The fiber $R$-module
structure on $T\Sol(M)$ given in \eqref{eqn:tangentVecSpstructure}
restricts to $T_{\sc}\Sol(M)$. As a consequence,
the generalized smooth space $\Gamma^{\infty}(T_{\sc}\Sol(M))$ of sections
of $\pi : T_{\sc}\Sol(M)\to \Sol(M)$ is an $R^{\Sol(M)}$-module and in particular 
an $R^{\Sol(M)}$-submodule of the module of vector fields $\Gamma^\infty(T\Sol(M))$ on $\Sol(M)$.
(See \eqref{eqn:Abgroup} and \eqref{eqn:RSolmodule} for the relevant module structure.) 
Finally, $\Gamma^\infty(T_{\sc}\Sol(M))$ is a Lie subalgebra
of $\Gamma^\infty(T\Sol(M))$, i.e.\ the Lie bracket of vector fields restricts
to $\Gamma^\infty(T_{\sc}\Sol(M))$,
\begin{flalign}\label{eqn:LieSC}
[\,\cdot\,,\,\cdot\,] : \Gamma^\infty(T_{\sc}\Sol(M))\times  \Gamma^\infty(T_{\sc}\Sol(M))\longrightarrow \Gamma^\infty(T_{\sc}\Sol(M))~.
\end{flalign}
This can be confirmed by using the explicit formula for the Lie bracket on $ \Gamma^\infty(T\Sol(M))$
that is given in Appendix \ref{app:veryexplicit}, see in particular \eqref{eqn:Liebracket}.

\paragraph{Presymplectic form:}
Restricting the global $\Omega^{m-1}(M)$-valued $1$-form $\theta_M^{}$
given in \eqref{eqn:thetaglobal} and \eqref{eqn:thetaglobalcomp} to  $ \Gamma^\infty(T_{\sc}\Sol(M))$,
it induces a $\CC$-morphism
\begin{flalign}
\theta_M^{} : \Gamma^\infty(T_{\sc}\Sol(M)) \longrightarrow \Omega^{m-1}_{\sc}(M)^{\Sol(M)}~,
\end{flalign}
where $\Omega^{m-1}_{\sc}(M)$ is the subsheaf of
$\Omega^{m-1}(M):\FMan^\op\to \Set$ that is specified by
\begin{multline}\label{eqn:scforms}
\Omega^{m-1}_{\sc}(M)(t) := 
\Big\{\omega \in \Omega^{m-1}(M)(t):\\
\forall x\in N~ \exists\, U \ni x \mbox{ open}\,,~ K\in \mathcal{K}_M^{}:~ 
\supp (\omega_U)\subseteq U \times J_M^{}(K) \Big\}~,
\end{multline}
for all objects $t$ in $\FMan$.  (Compare this with \eqref{eqn:Tscexplicit}.)
This claim can be easily confirmed using \eqref{eqn:thetaformcomp}.
In analogy to \eqref{eqn:currentOmegadd}, we define the presymplectic current
\begin{flalign}\label{eqn:ucurr}
\mathrm{u}_M^{} : \Gamma^\infty(T_{\sc}\Sol(M))\times \Gamma^\infty(T_{\sc}\Sol(M))\longrightarrow 
\Omega^{m-1}_{\sc,\,\dd}(M)^{\Sol(M)}
\end{flalign}
on generalized points $v,v^\prime : t \times \Sol(M)\to T_{\sc}\Sol(M)$, satisfying
the section condition $\pi\circ v^{(\prime)} = \pr_{\Sol(M)}^{}$, by Koszul's formula 
\begin{flalign}
\mathrm{u}_M^{}(v,v^\prime) := \dd \theta_M^{}(v,v^\prime) = v\big(\theta_M^{}(v^\prime)\big) - 
v^\prime\big(\theta_M^{}(v)\big) - \theta_M^{}\big([v,v^\prime]\big)~.
\end{flalign}
The explicit calculation performed in Appendix \ref{app:veryexplicit} 
is basically left unchanged by the restriction to spacelike compact vector fields. 
In particular, the formula for $\mathrm{u}_M^{}(v,v^\prime)$
provided in \eqref{eqn:presympcurrentultraexplicit} is still valid, 
however the result inherits the support restriction from the spacelike compact vector fields. 
Notice further that the same formula is still valid if we remove the support restriction 
on one of the arguments of $\mathrm{u}_M^{}$ and that this does not affect 
the spacelike compact support property of the differential forms $\Omega^{m-1}_{\sc,\,\dd}(M)$. 
In this way we obtain extensions (denoted by the same symbol)
\begin{subequations}\label{eqn:uextension}
\begin{flalign}
\mathrm{u}_M^{} : \Gamma^\infty(T_{\sc}\Sol(M)) \times \Gamma^\infty(T\Sol(M)) &\longrightarrow \Omega^{m-1}_{\sc,\,\dd}(M)^{\Sol(M)}~,\label{eqn:uextension1}\\
\mathrm{u}_M^{} : \Gamma^\infty(T\Sol(M)) \times \Gamma^\infty(T_{\sc}\Sol(M)) &\longrightarrow \Omega^{m-1}_{\sc,\,\dd}(M)^{\Sol(M)}~,\label{eqn:uextension2}
\end{flalign}
\end{subequations}
where only one factor is required to have spacelike compact support. 
We may now compose  \eqref{eqn:ucurr} with the integration $\CC$-morphism associated to a 
(not necessarily compact) Cauchy surface $\Sigma\hookrightarrow M$ and obtain a presymplectic form
\begin{flalign}\label{tmp:omegasc}
\omega_{M}^{} := {\int_{\Sigma}}^{\Sol(M)} \, \circ\,  \mathrm{u}_M^{} :  \Gamma^\infty(T_{\sc}\Sol(M))\times 
\Gamma^\infty(T_{\sc}\Sol(M)) 
\longrightarrow R^{\Sol(M)}~,
\end{flalign}
which does not depend on the choice of Cauchy surface. 
Composing the extensions \eqref{eqn:uextension} 
of $\mathrm{u}_M^{}$ with the integration $\CC$-morphism 
provides extensions (denoted by the same symbol) 
\begin{subequations}\label{eqn:omegaextension}
\begin{flalign}
\omega_{M}^{}  :  \Gamma^\infty(T_{\sc}\Sol(M))\times 
\Gamma^\infty(T\Sol(M))  &\longrightarrow R^{\Sol(M)}~,\label{eqn:omegaextension1}\\
\omega_{M}^{}  :  \Gamma^\infty(T\Sol(M))\times 
\Gamma^\infty(T_{\sc}\Sol(M))  &\longrightarrow R^{\Sol(M)}~,\label{eqn:omegaextension2}
\end{flalign}
\end{subequations}
where only one factor is required to have spacelike compact support. 
Notice that this support restriction is crucial in order to make sense of \eqref{eqn:omegaextension} 
as \eqref{tmp:omegasc} involves integration over a (possibly) non-compact Cauchy surface $\Sigma$.

\paragraph{Hamiltonian vector field equation and Poisson algebra:}
The extension \eqref{eqn:omegaextension1} of the presymplectic form 
can be adjoined to a $\CC$-morphism $\iota_{(\,\cdot\,)}(\omega_M^{}): \Gamma^\infty(T_\sc\Sol(M)) \to \Omega^1(\Sol(M))$,
which we use to define the generalized smooth space $\PP(M)$ as the pullback
\begin{flalign}\label{eqn:hamVFequationNC}
\xymatrix{
\ar@{-->}[d] \PP(M)\ar@{-->}[rr] && \Gamma^\infty(T_\sc\Sol(M))\ar[d]^-{\iota_{(\,\cdot\,)}(\omega_M^{})}\\
R^{\Sol(M)}\ar[rr]_-{\dd}&& \Omega^1(\Sol(M))
}
\end{flalign}
in $\CC$. The Poisson algebra structure which we developed in Section \ref{sec:compact}
for the case of compact Cauchy surfaces (cf.\ \eqref{eqn:PoisRalg} and \eqref{eqn:PoisPalg}) 
applies to the present case. In particular, all computations we performed to confirm the Poisson algebra properties
involve at most one non-spacelike compact vector field, so all occurring vector field insertions
into $\omega_M^{}$ are well-defined due to the extensions \eqref{eqn:omegaextension}.
Summing up, we obtain
\begin{theo}\label{theo:Poissongeneral}
Let $M$ be any object in $\Loc$. Then $\PP(M)$, defined as the pullback in \eqref{eqn:hamVFequationNC},
is a Poisson algebra object in the Cahiers topos $\CC$ when equipped with the
$R$-module structure inherited via pullback, the $R$-algebra structure \eqref{eqn:PoisRalg}
and the Poisson bracket \eqref{eqn:PoisPalg}.
\end{theo}

In the special case when $M$ admits a compact Cauchy surface $\Sigma\hookrightarrow M$,
the spacelike compact tangent bundle $T_\sc\Sol(M)$ coincides with
the full tangent bundle $T\Sol(M)$. In particular,
the Poisson algebra $\PP(M)$ given by Theorem \ref{theo:Poissoncompact}
coincides with the one given by Theorem \ref{theo:Poissongeneral}.

\begin{rem}
\footnote{We are grateful to the anonymous referee for the observation contained in this remark.} 
In \eqref{eqn:hamVFequationNC} we defined the vertical solid arrow by adjoining the extension 
\eqref{eqn:omegaextension1} of the presymplectic form. 
Similarly, we may as well adjoin the extension \eqref{eqn:omegaextension2} and consider the pullback
\begin{flalign}\label{eqn:PoisModdiag}
\xymatrix{
\ar@{-->}[d] \mathfrak{M}(M)\ar@{-->}[rr] && \Gamma^\infty(T\Sol(M))\ar[d]^-{\iota_{(\,\cdot\,)}(\omega_M^{})}\\
R^{\Sol(M)}\ar[r]_-{\dd}& \Omega^1(\Sol(M))\ar[r] &\Omega_\sc^1(\Sol(M))
}
\end{flalign}
in $\CC$. This is similar to \eqref{eqn:hamVFequationNC}, however without the requirement
that Hamiltonian vector fields are spacelike compact. As a consequence, their insertion 
in \eqref{eqn:omegaextension2} provides 1-forms $\Omega_\sc^1(\Sol(M))$ 
only defined with respect to the $R^{\Sol(M)}$-submodule $\Gamma^\infty(T_\sc\Sol(M))$ 
of $\Gamma^\infty(T\Sol(M))$ (the submodule inclusion induces the displayed $\CC$-morphism 
$\Omega^1(\Sol(M)) \to \Omega_\sc^1(\Sol(M))$). 
If $M$ does not admit compact Cauchy surfaces, then the formula in \eqref{eqn:PoisPalg} 
does {\em not} make sense on $\mathfrak{M}(M)$ because it involves an evaluation 
of the presymplectic form on two not necessarily spacelike compact vector fields.
However, we notice that $\mathfrak{M}(M)$ defined in \eqref{eqn:PoisModdiag} is a Poisson module \cite{Farkas}
over the Poisson algebra $\PP(M)$ of Theorem \ref{theo:Poissongeneral}.
The Poisson module structure is given by the explicit formulas \eqref{eqn:PoisRalg} and \eqref{eqn:PoisPalg}
after realizing that these are still well-defined when only one vector field is spacelike compact 
(cf.\ the extensions in \eqref{eqn:omegaextension}). 
Of course, $\mathfrak{M}(M)$ becomes a Poisson algebra canonically isomorphic to $\PP(M)$ 
whenever $M$ admits compact Cauchy surfaces. 
\end{rem}


\section{\label{sec:conclusion}Concluding remarks}
Our construction of the Poisson algebras $\PP(M)$ in 
Theorem \ref{theo:Poissongeneral} is rather abstract and in particular it
does not use any PDE-analytical properties of 
the field equation \eqref{eqn:EOMoperator} 
or its linearization \eqref{eqn:EOMoperatorlin}. 
The existence of Poisson algebras corresponding to non-linear field equations 
and their associated presymplectic currents is therefore a generic feature
of working in a topos theoretic setting. However, analyzing and proving additional
properties of the Poisson algebras, e.g.\ the validity of
the classical versions of the axioms of locally covariant quantum field theory
\cite{Brunetti:2001dx} including functoriality of the assignment $M\mapsto \PP(M)$,
requires a deeper analytical understanding of the field equation and its linearization.
\sk

One of the main tools available to analyze such additional properties
is the Cauchy problem of the field equation  \eqref{eqn:EOMoperator} 
and its linearization. We shall briefly explain how these are formalized
in our framework:
Given any object $M$ in $\Loc$ and any Cauchy surface $\Sigma$, 
with embedding denoted by $j: \Sigma\to M$, we define
a $\CC$-morphism
\begin{flalign}
\mathrm{data}_{\Sigma}^{} :=\big(R^{j}, \ast_{\Sigma}^{}\circ \Omega^{m-1}(j)\circ \ast_M^{}\circ \dd_M^{}\big) :  R^M\longrightarrow R^\Sigma\times R^\Sigma~,
\end{flalign}
which assigns to a field configuration its ``initial position'' and ``initial velocity'' on $\Sigma$.
Denoting the $\CC$-morphisms $R^{j} : R^M \to R^\Sigma$ and $\Omega^{m-1}(j) : \Omega^{m-1}(M)\to 
\Omega^{m-1}(\Sigma)$ for notational simplicity by $j^\ast$, the components 
of this natural transformation are given by
\begin{flalign}
\mathrm{data}_{\Sigma}^{}  : R^M(t)\longrightarrow R^\Sigma(t) \times R^\Sigma(t)~,~~
\Phi \longmapsto \big( j^\ast (\Phi)  , \ast_{\Sigma}^{} \,j^\ast (\ast_M^{}\dd_M^{}\Phi)\big)~.
\end{flalign}
We say that the Cauchy problem of the non-linear field equation \eqref{eqn:EOMoperator} is well-posed
on $M$ if 
\begin{flalign}
\mathrm{data}_\Sigma^{} : \Sol(M)\longrightarrow R^\Sigma\times R^\Sigma~
\end{flalign}
is an isomorphism in $\CC$, for any Cauchy surface $j :\Sigma \to M$.
\sk

Well-posedness of the Cauchy problem in our framework is thus equivalent
to solve the field equation $P_M^{}(\Phi)=0$ for $\Phi \in C^\infty(N\times M)\otimes_{\bbR} W$
for any given initial datum $\phi,\pi \in C^\infty(N\times \Sigma)\otimes_{\bbR} W$. 
Notice that these are Cauchy problems on $M$, where the field equation and initial conditions
are smoothly parametrized by $N$ and $W$. As a consequence, the solution theory of such equations
requires a detailed understanding of the smooth
parameter and initial-value dependence of solutions to the 
ordinary field equation $P_M^{}(\Phi) =0$, for $\Phi \in C^\infty(M)$.
\sk

The linearized Cauchy problem can be formalized as follows:
Consider the commutative diagram
\begin{flalign}\label{eqn:pullbackbundle}
\xymatrix{
T\Sol(M) \ar@{-->}[rrd]^-{\exists !} \ar@/_1pc/[ddrr]_-{\pi}\ar@/^1pc/[rrrrd]^-{T\mathrm{data}_{\Sigma}^{}}&& && \\
&& \ar@{-->}[d]\mathrm{data}_\Sigma^\ast T(R^\Sigma\times R^\Sigma) \ar@{-->}[rr] && T(R^\Sigma\times R^\Sigma)\ar[d]^-{\pi}\\
&& \Sol(M) \ar[rr]_-{\mathrm{data}_\Sigma^{}} && R^\Sigma \times R^\Sigma
}
\end{flalign}
in $\CC$, where $\mathrm{data}_\Sigma^\ast T(R^\Sigma\times R^\Sigma)\to \Sol(M)$ 
is the pullback along $\mathrm{data}_{\Sigma}^{}$
of the tangent bundle of the generalized smooth space of initial data $R^\Sigma\times R^\Sigma$. The
unique $\CC$-morphism to $\mathrm{data}_\Sigma^\ast T(R^\Sigma\times R^\Sigma)$ that is 
depicted in the diagram \eqref{eqn:pullbackbundle} has components given by 
\begin{flalign}
 T\Sol(M)(t)\longrightarrow  \mathrm{data}_\Sigma^\ast T(R^\Sigma\times R^\Sigma)(t)~,~~
\Phi + \epsilon\,\Psi \longmapsto  \big( \Phi , \mathrm{data}_{\Sigma}^{}(\Phi) + \epsilon\,\mathrm{data}_{\Sigma}^{} (\Psi) \big)~.
\end{flalign}
We say that the {\em linearized} Cauchy problem of the
non-linear field equation \eqref{eqn:EOMoperator} is well-posed on $M$ if this arrow
is an isomorphism in $\CC$, for any Cauchy surface $j :\Sigma \to M$.
\sk

Well-posedness of the linearized Cauchy problem can be
formulated equivalently  in the following more elementary terms:
Given any generalized point $\Phi: t\to \Sol(M)$,
then there exists for each generalized point $(\psi,\chi) : t\to R^\Sigma\times R^\Sigma$
a unique $\Psi \in C^\infty(N\times M) \otimes_{\bbR} W$ which satisfies the linearized equation
of motion $P^\mathrm{lin}_{M,\,\Phi}\Psi =0$ around $\Phi$ (cf.\ \eqref{eqn:EOMoperatorlin})
and the initial condition $\mathrm{data}_{\Sigma}^{}(\Psi) = (\psi,\chi)$.
Again, these are Cauchy problems on $M$, where the field equation and initial conditions
are smoothly parametrized by $N$ and $W$.
\sk

It is easy to see that if the Cauchy problem is well-posed, then also the linearized Cauchy problem is well-posed:
If $\mathrm{data}_{\Sigma}^{}$ is an isomorphism, then so is $T\mathrm{data}_{\Sigma}^{}$.
The claim then follows from $ \mathrm{data}_\Sigma^\ast T(R^\Sigma\times R^\Sigma)
\simeq T(R^\Sigma\times R^\Sigma)$ and the commutative diagram \eqref{eqn:pullbackbundle}.
\sk

A detailed investigation of linearized and in particular non-linear 
Cauchy problems in our framework 
is beyond the scope of this paper and will be 
addressed in a future work. 
There we will also attempt to confirm the classical versions of the axioms of 
locally covariant quantum field theory \cite{Brunetti:2001dx} for simple examples of 
non-linear field equations \eqref{eqn:EOMoperator}.


\section*{Acknowledgments}
We would like to thank Klaus Fredenhagen for useful discussions and comments, 
as well as the anonymous referees, whose suggestions helped us improving the quality of the paper 
and encouraged us to further emphasize the advantages of an approach based on synthetic differential geometry.
We also would like to thank the Mathematisches Forschungsinstitut Oberwolfach (MFO) 
for the great hospitality during the ``Derived Geometry Seminar'',
where part of this work has been performed.
The work of M.B.\ is supported by a Postdoctoral Fellowship of the Alexander von Humboldt Foundation (Germany). 
The work of A.S.\ was supported by a Research Fellowship of the Deutsche Forschungsgemeinschaft (DFG, Germany).


\appendix

\section{\label{app:veryexplicit}Explicit expression for the presymplectic current}
In this appendix we compute explicitly the presymplectic current \eqref{eqn:presympcurrent}.
This result is needed to confirm that $\mathrm{u}_M^{}$ 
takes values in the generalized smooth space $\Omega^{m-1}_{\dd}(M)^{\Sol(M)}$.
\sk

Let 
\begin{subequations}\label{eqn:vecexplicit}
\begin{flalign}
v : t\times\Sol(M)\longrightarrow T\Sol(M)
\end{flalign} 
be a generalized point of $\Gamma^\infty(T\Sol(M))$,
i.e.\ a $\CC$-morphism which satisfies the section condition $\pi \circ  v= \pr_{\Sol(M)}^{}$.
The components of $v$ are maps of sets
\begin{flalign}
v_{t^\prime} : \Hom_{\FMan}(t^\prime,t)\times \Sol(M)(t^\prime) \longrightarrow T\Sol(M)(t^\prime)~,~~
 (f , \Phi )& \longmapsto \Phi + \epsilon\,\Psi_{t^\prime}(f,\Phi)~,
\end{flalign}
\end{subequations}
where the tangent vectors $\Psi_{t^\prime}$ depend on $\FMan$-morphisms $f : t^\prime \to t$
and the base point $\Phi$.
\sk

Recalling that $T\Sol(M) = \Sol(M)^D$ is an exponential object,
we can adjoin $D$ and equivalently regard
the generalized point $v : t\times\Sol(M)\to T\Sol(M)$ 
as a $\CC$-morphism
\begin{subequations}\label{eqn:vecexplicitalternative}
\begin{flalign}
\widetilde{v} : t\times \Sol(M)\times D \longrightarrow \Sol(M)~.
\end{flalign}
By a short calculation involving Yoneda's Lemma, we obtain that the components
of $\widetilde{v}$ are given by
\begin{flalign}
\nn \widetilde{v}_{t^\prime} : \Hom_{\FMan}(t^\prime,t) \times \Sol(M)(t^\prime)\times\Hom_{\FMan}(t^\prime, D) &\longrightarrow
\Sol(M)(t^\prime)~,\\
(f,\Phi, \delta) &\longmapsto \Phi + \delta(\epsilon)\,\Psi_{t^\prime}(f,\Phi)~,
\end{flalign}
\end{subequations}
where on the right-hand-side we have used that the $\FMan$-morphism $\delta : t^\prime\to D$
is by definition a $C^\infty$-ring morphism $\delta : \bbR[\epsilon] \to C^\infty(N^\prime)\otimes_{\bbR} W^\prime$,
where $t^\prime = N^\prime \times\ell W^\prime$.
\sk

Next, we address the action of vector fields as directional derivatives on the generalized
smooth space $\Omega^{m-1}(M)^{\Sol(M)}$ of $\Omega^{m-1}(M)$-valued functions on $\Sol(M)$.
A generalized point of $\Omega^{m-1}(M)^{\Sol(M)}$ is a $\CC$-morphism
\begin{subequations}\label{eqn:functionexplicit}
\begin{flalign}
\zeta : t\times \Sol(M)\longrightarrow \Omega^{m-1}(M)~,
\end{flalign}
and we shall denote its components by
\begin{flalign}
\zeta_{t^\prime} : \Hom_{\FMan}(t^\prime,t)\times \Sol(M)(t^\prime) \longrightarrow \Omega^{m-1}(M)(t^\prime)~,~~
(f,\Phi)  \longmapsto \zeta_{t^\prime}(f,\Phi)~,
\end{flalign}
\end{subequations}
where according to \eqref{eqn:formspace} $ \zeta_{t^\prime}(f,\Phi)$ is an element
in $\Omega^{0,m-1}(N^\prime\times M)\otimes_{\bbR} W^\prime$, 
for $t^\prime = N^\prime\times\ell W^\prime$.
We can compose \eqref{eqn:functionexplicit} with \eqref{eqn:vecexplicitalternative}
according to the commutative diagram
\begin{subequations}\label{eqn:functionveccomposition}
\begin{flalign}
\xymatrix{
\ar[d]_-{(\pr_{t}^{}, \widetilde{v})}t\times\Sol(M)\times D \ar[rr]^-{\zeta\bullet \widetilde{v}} && \Omega^{m-1}(M)\\
t\times \Sol(M) \ar[rru]_-{\zeta}&&
}
\end{flalign}
in $\CC$. Explicitly, the components of $\zeta\bullet \widetilde{v}$ are given by
\begin{flalign}
\nn(\zeta\bullet \widetilde{v})_{t^\prime} : \Hom_{\FMan}(t^\prime,t)\times \Sol(M)(t^\prime)\times \Hom_{\FMan}(t^\prime,D)
&\longrightarrow \Omega^{m-1}(M)(t^\prime)~,\\
(f, \Phi,\delta)& \longmapsto \zeta_{t^\prime}\big(f, \Phi + \delta(\epsilon)\,\Psi_{t^\prime}(f,\Phi)\big)~.
\end{flalign}
\end{subequations}
Using naturality of the components of $\zeta$, we observe that
\begin{flalign}\label{eqn:dertmp1}
\zeta_{t^\prime}\big(f, \Phi + \delta(\epsilon)\,\Psi_{t^\prime}(f,\Phi)\big) = 
 \delta\big(\zeta_{t^\prime\times D }\big( f\circ \pr_{t^\prime}^{}, \Phi + \epsilon\,\Psi_{t^\prime}(f,\Phi)\big)\big)~.
\end{flalign}
More explicitly, we expand
$\zeta_{t^\prime\times D }( f\circ \pr_{t^\prime}^{}, \Phi + \epsilon\,\Psi_{t^\prime}(f,\Phi))\in 
\Omega^{0,m-1}(N^\prime \times M)\otimes_{\bbR}W^\prime\otimes_{\bbR}\bbR[\epsilon]$
in terms of $\epsilon$ as
\begin{flalign}\label{eqn:dertmp2}
\zeta_{t^\prime\times D }\big( f\circ \pr_{t^\prime}^{}, \Phi + \epsilon\,\Psi_{t^\prime}(f,\Phi)\big) = 
\zeta_{t^\prime}\big(f,\Phi\big) + \epsilon \, v(\zeta)_{t^\prime} \big(f,\Phi\big)~,
\end{flalign}
where the explicit form of the $\epsilon^0$-term follows from the section condition of $v$
and the $\epsilon^1$-term is defined by this expansion.
Then \eqref{eqn:functionveccomposition} is given by
\begin{flalign}
\zeta_{t^\prime}\big(f, \Phi + \delta(\epsilon)\,\Psi_{t^\prime}(f,\Phi)\big) 
= \zeta_{t^\prime}\big(f,\Phi\big) + \delta(\epsilon) \, v(\zeta)_{t^\prime} \big(f,\Phi \big)~,
\end{flalign}
and we define the directional derivative
\begin{subequations}\label{eqn:directionalder}
\begin{flalign}
v(\zeta) : t\times \Sol(M) \longrightarrow \Omega^{m-1}(M)
\end{flalign}
of $\zeta$ along $v$ by the components
\begin{flalign}
v(\zeta)_{t^\prime} : \Hom_{\FMan}(t^\prime,t)\times\Sol(M)(t^\prime) \longrightarrow \Omega^{m-1}(M)(t^\prime)~,~~
(f,\Phi) \longmapsto v(\zeta)_{t^\prime} \big(f,\Phi\big)~.
\end{flalign}
\end{subequations}

We can now compute the first term $v(\theta_M^{}(v^\prime))$ of the presymplectic current
\eqref{eqn:presympcurrent}. The components of
\begin{subequations}\label{eqn:thetavprime}
\begin{flalign}
\theta_M^{}(v^\prime) : t\times \Sol(M) \longrightarrow \Omega^{m-1}(M)
\end{flalign}
can be easily computed by \eqref{eqn:thetaformcomp} and are given by
\begin{flalign}
\nn \theta_M^{}(v^\prime)_{t^\prime} : \Hom_{\FMan}(t^\prime,t) \times \Sol(M)(t^\prime) &\longrightarrow \Omega^{m-1}(M)(t^\prime)~,\\
(f,\Phi)&\longmapsto -\Psi_{t^\prime}^\prime(f,\Phi) \,\ast_M^{}\dd_M^{}\Phi~,
\end{flalign}
\end{subequations}
where we have used a notation similar to \eqref{eqn:vecexplicit} for the components 
$v^\prime_{t^\prime} : (f,\Phi)\mapsto \Phi  +\epsilon \,\Psi_{t^\prime}^{\prime}(f,\Phi)$ of $v^\prime$. 
Motivated by \eqref{eqn:dertmp1} we compute
\begin{flalign}
\nn &\theta_M^{}(v^\prime)_{t^\prime\times D} \big(f\circ \pr_{t^\prime}^{} , \Phi + \epsilon \,\Psi_{t^\prime}(f,\Phi) \big)\\
\nn &\qquad~\qquad = -\Psi_{t^\prime\times D}^{\prime}\big(f\circ\pr_{t^\prime}^{},\Phi + \epsilon \,\Psi_{t^\prime}(f,\Phi)\big)\,\ast_M^{}\dd_M^{}
\big(\Phi + \epsilon \,\Psi_{t^\prime}(f,\Phi)\big) \\
&\qquad~\qquad = -\big( \Psi^\prime_{t^\prime}(f,\Phi) + \epsilon\, v(\Psi^\prime)_{t^\prime}(f,\Phi)\big)\, \ast_M^{}\dd_M^{}\big(\Phi + \epsilon \,\Psi_{t^\prime}(f,\Phi)\big)~,
\end{flalign}
where in the last equality we have used a notation for $\Psi^\prime$ similar to 
the one used in \eqref{eqn:dertmp2} for $\zeta$ in order to denote the expansion in $\epsilon$.
Using \eqref{eqn:directionalder}, we thus obtain that 
the components of the $\CC$-morphism
\begin{subequations}\label{eqn:Koszul1}
\begin{flalign}
v\big(\theta_M^{}(v^\prime)\big) : t\times\Sol(M) \longrightarrow \Omega^{m-1}(M)
\end{flalign}
are given by
\begin{flalign}
\nn v\big(\theta_M^{}(v^\prime)\big)_{t^\prime} : \,&\Hom_{\FMan}(t^\prime,t) \times \Sol(M)(t^\prime) \longrightarrow \Omega^{m-1}(M)(t^\prime)~,\\
(f,\Phi)&\longmapsto - \Psi^\prime_{t^\prime}(f,\Phi) \, \ast_M^{}\dd_M^{} \Psi_{t^\prime}(f,\Phi)
-  v(\Psi^\prime)_{t^\prime}(f,\Phi)\, \ast_M^{}\dd_M^{} \Phi~.
\end{flalign}
\end{subequations}
Simply exchanging $v$ and $v^\prime$ one gets also the second term in the presymplectic current \eqref{eqn:presympcurrent}.
\sk

It remains to understand the Lie bracket 
\begin{flalign}
[\,\cdot\,,\,\cdot\,] : \Gamma^\infty(T\Sol(M))\times \Gamma^\infty(T\Sol(M)) \longrightarrow \Gamma^\infty(T\Sol(M))
\end{flalign}
involved in the definition of the presymplectic current \eqref{eqn:presympcurrent}.
Given two generalized points $v,v^\prime : t\to \Gamma^\infty(T\Sol(M))$,
we regard them as in \eqref{eqn:vecexplicitalternative} as $\CC$-morphisms
$\widetilde{v},\widetilde{v^\prime} : t\times \Sol(M)\times D\to \Sol(M)$.
Following \cite[Chapter 3.2.2]{Lavendhomme}, we define a $\CC$-morphism (depending on $v$ and $v^\prime$)
\begin{subequations}\label{eqn:taumap}
\begin{flalign}
\tau : t\times\Sol(M) \times D\times D \longrightarrow \Sol(M)~,
\end{flalign}
by setting for its components
\begin{flalign}
\nn \tau_{t^\prime} : \Hom_{\FMan}(t^\prime,t)& \times \Sol(M)(t^\prime) \times \Hom_{\FMan}(t^\prime,D)\times \Hom_{\FMan}(t^\prime, D) \longrightarrow \Sol(M)(t^\prime)~,\\
(f,\Phi,\delta_1,\delta_2)&\longmapsto \widetilde{v^\prime}_{t^\prime}\left(f,  \widetilde{v}_{t^\prime}\left(f,  \widetilde{v^\prime}_{t^\prime}\left( f,  \widetilde{v}_{t^\prime}\big(f,\Phi, \delta_1\big), \delta_2  \right), \overline{\delta_1}\right) , \overline{\delta_2}\right)~,
\end{flalign}
\end{subequations}
where $\overline{\delta} : \bbR[\epsilon] \to C^\infty(N^\prime)\otimes_{\bbR} W^\prime\,,~a + \epsilon\,b\mapsto
a - \delta(\epsilon)\,b$ is the $C^\infty$-ring morphism induced by $\delta :  \bbR[\epsilon] \to C^\infty(N^\prime)
\otimes_{\bbR} W^\prime\,,~a +\epsilon \,b\mapsto a + \delta(\epsilon)\,b$ and flipping the sign in front of
$\epsilon$.
Using \eqref{eqn:vecexplicitalternative} and arguments similar to \eqref{eqn:dertmp1} and \eqref{eqn:dertmp2},
we can expand the components of $\tau$ in terms of $\delta_1(\epsilon)$ and $\delta_2(\epsilon)$.
Using that $\delta_1(\epsilon)^2 =\delta_1(\epsilon^2)=0$ and similarly $\delta_2(\epsilon)^2 =\delta_2(\epsilon^2)=0$,
this expansion stops at order $\delta_1(\epsilon)\,\delta_2(\epsilon)$. 
The component $[v,v^\prime ]_{t^\prime}$ of the Lie bracket is then defined by setting 
$\delta =\delta_1 \cdot \delta_2: \bbR[\epsilon] \to C^\infty(N^\prime)\otimes_{\bbR} W^\prime$ 
and then going back from \eqref{eqn:vecexplicitalternative} to \eqref{eqn:vecexplicit}.
We explicitly obtain that the components of the Lie bracket
\begin{subequations}\label{eqn:Liebracket}
\begin{flalign}
[v,v^\prime ] : t\times\Sol(M) \longrightarrow T\Sol(M)~,
\end{flalign}
are given by
\begin{flalign}
\nn [v,v^\prime ]_{t^\prime} : \Hom_{\FMan}(t^\prime,t) \times\Sol(M)(t^\prime) &\longrightarrow T\Sol(M)(t^\prime)~,\\
(f,\Phi) &\longmapsto \Phi + \epsilon\, \Big(v(\Psi^\prime)_{t^\prime}(f,\Phi) - v^\prime(\Psi)_{t^\prime}(f,\Phi)\Big)~,
\end{flalign}
\end{subequations}
where  $v(\Psi^\prime)$ and $v^\prime(\Psi)$ are defined as in \eqref{eqn:Koszul1}. 
With reference to \eqref{eqn:LieSC}, we stress that, 
for $v,v^\prime$ generalized points of $\Gamma^\infty(T_{\sc}\Sol(M))$, 
the restriction on the support is preserved by the Lie bracket. 
This fact can be directly read off from the formula 
and it means that the Lie bracket closes on $\Gamma^\infty(T_{\sc}\Sol(M))$.
\sk

Combining \eqref{eqn:Koszul1} and \eqref{eqn:Liebracket}, we obtain the following expression for the components
of the presymplectic current \eqref{eqn:presympcurrent}:
\begin{flalign}\label{eqn:presympcurrentultraexplicit}
\nn \mathrm{u}_M^{}(v,v^\prime)_{t^\prime} :\,& \Hom_{\FMan}(t^\prime,t)\times\Sol(M)(t^\prime)\longrightarrow \Omega^{m-1}(M)(t^\prime)~,\\
(f,\Phi)& \longmapsto  \Psi_{t^\prime}(f,\Phi) \, \ast_M^{}\dd_M^{} \Psi_{t^\prime}^{\prime}(f,\Phi) 
- \Psi^\prime_{t^\prime}(f,\Phi) \, \ast_M^{}\dd_M^{} \Psi_{t^\prime}(f,\Phi)~.
\end{flalign}
Applying $\dd_M^{}$ on this expression and recalling that
both $\Psi^\prime_{t^\prime}(f,\Phi) $ and $\Psi_{t^\prime}(f,\Phi) $ satisfy the
linearized field equation \eqref{eqn:EOMoperatorlin} around
the same $\Phi$, it is easy to confirm that $\mathrm{u}_M^{}(v,v^\prime)$ 
is a $\CC$-morphism to the generalized smooth space of closed forms $\Omega^{m-1}_{\dd}(M)$.
For completeness, we add below the relevant calculation 
using a compact notation where all indices and arguments are omitted: 
\begin{flalign}
\nn \dd\big(\Psi\,\ast\dd \Psi^\prime - \Psi^\prime \,\ast\dd \Psi \big)
&= \dd\Psi\wedge \ast \dd\Psi^\prime + \Psi \,\dd \ast\dd \Psi^\prime - 
\dd\Psi^\prime \wedge \ast\dd \Psi - \Psi^\prime \,\dd\ast\dd \Psi\\
&= \big(\Psi \,\square\Psi^\prime  - \Psi^\prime\,\square\Psi\big)\,\vol= -\big(\Psi\, \rho^\prime(\Phi)\,\Psi^\prime  - \Psi^\prime\,\rho^\prime(\Phi)\,\Psi\big)\,\vol =0~.
\end{flalign}
In the first step we used the Leibniz rule and in the second step the
linearized equation of motion $\square \Psi^{(\prime)} + \rho^\prime(\Phi)\, \Psi^{(\prime)} =0$
and the property $\dd\Psi\wedge \ast \dd\Psi^\prime = \dd\Psi^\prime\wedge \ast \dd\Psi$ 
of the Hodge operator.

\end{document}